\documentclass[12pt,usenames,dvipsnames]{article}

\usepackage{latexsym}
\usepackage{amssymb,amsfonts,amsmath}
\usepackage{graphicx} 
\usepackage{indentfirst}
\usepackage{bbm}
\usepackage{amssymb}
\usepackage{verbatim}
\usepackage{amsmath, amsthm,amssymb}
\usepackage{mathrsfs}
\usepackage{hyperref}
\usepackage{amsfonts}
\usepackage{dsfont}
\usepackage{cite}
\usepackage{xcolor}
\usepackage{enumerate}
\usepackage{cleveref}

\parskip=\medskipamount

\newcommand {\cD}{{\cal D}}
\newcommand {\cE}{{\cal E}}

\newcommand {\cJ}{{\cal J}}

\newcommand {\cM}{{\cal M}}
\newcommand {\cN}{{\cal N}}
\newcommand {\cO}{{\cal O}}
\newcommand {\cP}{{\cal P}}

\newcommand {\cS}{{\cal S}}

\newcommand {\cU}{{\cal U}}

\def\a{\alpha}

\def\b{\beta}

\def\d{\delta}
\def\e{\epsilon}

\def\g{\gamma}
\def\G{\Gamma}

\def\k{\kappa}
\def\l{\lambda}
\def\m{\mu}
\def\n{\nu}
\def\o{\omega}
\def\p{\pi}
\def\q{\theta}
\def\r{\rho}
\def\s{\sigma}
\def\t{\tau}

\def\x{\xi}

\def\F{\Phi}

\def\L{\Lambda}
\def\O{\Omega}

\def\Q{\Theta}
\def\S{\Sigma}

\def\rd{{\rm d}}
\def\ri{{\rm i}}
\def\re{{\rm e}}

\def\N{{\cal N}}

\newcommand{\ad}{{\dot{\alpha}}}                           
\newcommand{\bd}{{\dot{\beta}}}                            
\newcommand{\ve}{\varepsilon}                            
\newcommand{\cDB}{{\bar\cD}}                            

\newcommand{\ab}{{\a\b}}

\renewcommand{\aa}{{\a\ad}}
\newcommand{\bb}{{\b\bd}}
\newcommand{\pa}{\partial}                           
\newcommand{\hf}{\frac12}

\newcommand{\be}{\begin{equation}}
\newcommand{\ee}{\end{equation}}
\newcommand{\bea}{\begin{eqnarray}}
\newcommand{\eea}{\end{eqnarray}}
\newcommand{\non}{\nonumber}
\newcommand{\bm}[1]{\mbox{\boldmath$#1$}}

\def\double #1{#1{\hbox{\kern-2pt $#1$}}}

\newcommand{\hal}{{\hat{\a}}}
\newcommand{\hbe}{{\hat{\b}}}

\newcommand{\gd}{{\dot\g}}
\newcommand{\dd}{{\dot\d}}

\newcommand{\ts}{{\tilde{\s}}}

\renewcommand{\Bar}{\overline}

\newif\ifdtup

\newcommand{\bsubeq}{\begin{subequations}}
\newcommand{\esubeq}{\end{subequations}}

\newcommand{\eol}{\notag \\}

\numberwithin{equation}{section}

\newcommand{\sSp}{\mathsf{Sp}}
\newcommand{\sSU}{\mathsf{SU}}
\newcommand{\sSL}{\mathsf{SL}}
\newcommand{\sGL}{\mathsf{GL}}
\newcommand{\sSO}{\mathsf{SO}}
\newcommand{\sO}{\mathsf{O}}
\newcommand{\sU}{\mathsf{U}}

\newcommand{\sOSp}{\mathsf{OSp}}
\newcommand{\sSpin}{\mathsf{Spin}}
\newcommand{\sMat}{\mathsf{Mat}}

\newcommand{\me}{\hat{\ve}}

\newcommand{\dmu}{{\dot{\mu}}}
\newcommand{\dnu}{{\dot{\nu}}}

\newcommand{\T}{\text{T}}
\newcommand{\sT}{\text{sT}}
\newcommand{\braket}[2]{\langle #1 | #2 \rangle}

\begin{document}

\begin{titlepage}
\begin{flushright}
August, 2023
\end{flushright}
\vspace{5mm}

\begin{center}
{\Large \bf 
Embedding formalism for ${\cal N}$-extended AdS superspace in four dimensions
}
\end{center}

\begin{center}

{\bf Nowar E. Koning, Sergei M. Kuzenko and Emmanouil S. N. Raptakis} \\
\vspace{5mm}

\footnotesize{ 
{\it Department of Physics M013, The University of Western Australia\\
35 Stirling Highway, Perth W.A. 6009, Australia}}  
~\\
\vspace{2mm}
~\\
Email: \texttt{nowar.koning@research.uwa.edu.au,
sergei.kuzenko@uwa.edu.au, emmanouil.raptakis@uwa.edu.au }\\
\vspace{2mm}

\end{center}

\begin{abstract}
\baselineskip=14pt
 The supertwistor and bi-supertwistor formulations for ${\cal N}$-extended anti-de Sitter (AdS) superspace in four dimensions, ${\rm AdS}^{4|4\cal N}$, were derived two years ago in \cite{KT-M21}. In the present paper, we introduce a novel realisation of the ${\cal N}$-extended AdS supergroup $\mathsf{OSp}(\mathcal{N}|4;\mathbb{R})$ and 
apply it to develop a coset construction for ${\rm AdS}^{4|4\cal N}$ and the corresponding differential geometry.
This realisation naturally leads to an atlas on ${\rm AdS}^{4|4\cal N}$ (that is a generalisation of the stereographic projection for a sphere) that consists of two charts with chiral transition functions for ${\cal N}>0$. A manifestly  $\mathsf{OSp}(\mathcal{N}|4;\mathbb{R})$ invariant model for a superparticle in ${\rm AdS}^{4|4\cal N}$ is proposed. Additionally, by employing a conformal superspace approach, we describe the most general conformally flat $\cN$-extended supergeometry. This construction is then specialised to the case of ${\rm AdS}^{4|4\cal N}$.
\end{abstract}
\vspace{5mm}

\vfill

\vfill
\end{titlepage}

\newpage
\renewcommand{\thefootnote}{\arabic{footnote}}
\setcounter{footnote}{0}

\tableofcontents{}
\vspace{1cm}
\bigskip\hrule

\allowdisplaybreaks

\section{Introduction}

The simplest AdS superspace in four dimensions, $\rm{AdS}^{4|4} $,
was introduced in the early years of supersymmetry by Keck \cite{Keck} and Zumino \cite{Zumino77} as the coset superspace\footnote{We remind the reader that $\sSpin(3, 1 ) \cong \sSL (2,{\mathbb C})$ is the double covering group  of the connected component of the Lorentz group, $\sSO_0(3,1) \cong \sSpin(3,1 ) / {\mathbb Z}_2$.}
\bea
\text{AdS}^{4|4} = \frac{\sOSp(1|4;\mathbb{R})}{\sSpin(3,1)}\,,
\label{1.1}
\eea
and the thorough study of general superfield representations on $\rm{AdS}^{4|4} $ was given by Ivanov and Sorin \cite{IS}. It was also realised that  $\rm{AdS}^{4|4} $ originates as a maximally supersymmetric solution in the following off-shell formulations for $\cN=1$ supergravity:
(i)  the old minimal supergravity
\cite{Siegel77-77,WZ,old1,old2}
with a cosmological term \cite{Townsend}, 
see \cite{GGRS,BK} for a review; and (ii) the non-minimal AdS supergravity \cite{BK11}.

The group-theoretic realisation \eqref{1.1} of $\cN=1$ AdS superspace has a natural extension to the $\cN > 1$ case (see, e.g., \cite{Castellani})
\bea
\text{AdS}^{4|4\cN} = \frac{\sOSp(\cN|4;\mathbb{R})}
{ \sSpin(3, 1 ) \times \sO(\cN) }\,.
\label{1.2}
\eea
The description of $\rm{AdS}^{4|8} $ as  a maximally supersymmetric solution in  the 
minimal off-shell formulation for $\cN=2$ supergravity 
with a cosmological term, developed by  de Wit, Philippe and Van Proeyen  \cite{deWPV},
was given in \cite{KLRT-M1,KT-M08,Butter:2011ym}.\footnote{Pure $\cN=2$ supergravity in four dimensions was constructed by Ferrara and van Nieuwenhuizen in 1976 \cite{FvN}, and pure $\cN=2$ supergravity with a cosmological term was constructed by Freedman and Das \cite{FD}.} 

The conformal flatness of AdS$^{4|4}$ was first established in \cite{IS}, and it was later re-derived in textbooks \cite{GGRS,BK} within the supergravity framework. 
The superconformal flatness of $\rm{AdS}^{4|4\cN}$ was demonstrated in \cite{BILS} for arbitrary $\cN$.
Alternative proofs of the conformal flatness of $\text{AdS}^{4|8} $
were given in \cite{KT-M08,BK2011} using the off-shell $\cN=2$ supergravity framework. Ref. \cite{BKLT-M} described different  conformally flat realisations for  AdS$^{4|4}$  and AdS$^{4|8}$ which are based on the use of Poincar\'e coordinates.

In the non-supersymmetric case, there exist two different 
but equivalent realisations of AdS$_d$:
(i) as the coset space $\sO(d-1, 2)/\sO(d-1,1)$; and (ii) as 
a hypersurface in 
${\mathbb R}^{d-1,2}$ 
\bea
 -(Z^0)^2 + (Z^1)^2 + \dots +(Z^{d-1})^2 - (Z^d)^2 
= -\ell^2 ={\rm const}~.
 \label{Embedding0}
\eea
Both realisations of AdS$_d$ have found numerous applications in the literature. 
As regards $\text{AdS}^{4|4\cN}$, only the coset superspace realisation \eqref{1.2} 
had existed for many years. 
The supertwistor and bi-supertwistor formulations for  ${\rm AdS}^{4|4\cal N}$ have recently been developed \cite{KT-M21}. 
Analogous results in three dimensions have been derived in \cite{KT-M21,KT}.
In this paper we elaborate on the superembedding formalism\footnote{For a pedagogical review of superembeddings see \cite{Sorokin,Bandos:2023web}.} 
for ${\rm AdS}^{4|4\cal N}$.

Since the work by Ferber \cite{Ferber}, supertwistors have found numerous applications in  theoretical and mathematical physics. In particular, 
supertwistor realisations of compactified  $\cN$-extended Minkowski superspaces 
have been developed  in four
\cite{Manin,KNiederle} and three  \cite{Howe:1994ms,KPT-MvU}  dimensions, and their harmonic/projective extensions  have been derived
\cite{Rosly2,LN,Howe:1994ms,HH1,HH2,K-compactified06, K-compactified12,KPT-MvU,BKS}.\footnote{Similar ideas  were applied in Ref. \cite{Kuzenko:2014yia} to develop
supertwistor realisations of the  $2n$-extended supersphere $S^{3|4n}$,
with $n=1,2,\dots$, as a homogeneous space of the three-dimensional
Euclidean superconformal group $\sOSp(2n|2,2)$.}
Recently, supertwistor formulations for conformal supergravity theories in diverse dimensions have been proposed \cite{HL1,HL2}.
To the best of our knowledge, the supertwistor realisations of AdS superspaces in three and four dimensions have been given only in \cite{KT-M21,KT}, although 
(super)twistor descriptions of (super)particles in AdS spaces  had been studied in the literature earlier 
\cite{CGKRZ,CRZ,CKR,BLPS,Zunger,Cederwall1,Cederwall2,AB-GT1,AB-GT2,Uvarov} (see also \cite{ASW}).\footnote{We are grateful to Alex Arvanitakis for 
bringing Ref. \cite{ASW} to our attention.}

This paper is organised as follows. In section \ref{section2} we give a brief review of the (bi)supertwistor description of AdS$^{4|4\N}$ and present a manifestly  $\mathsf{OSp}(\mathcal{N}|4;\mathbb{R})$ invariant model for a superparticle in ${\rm AdS}^{4|4\cal N}$.
Section \ref{section3} is devoted to presenting a novel realisation of the AdS supergroup, which is then used in section \ref{section4} to develop a coset construction for ${\rm AdS}^{4|4\cal N}$. The coset construction is applied in section \ref{section5} to work out the differential geometry of ${\rm AdS}^{4|4\cal N}$. 
In section \ref{section6}, by employing the framework of conformal superspace, we describe the most general conformally flat supergeometry and then specialise our construction to the case of $\text{AdS}^{4|4\cN}$.
The main body of the paper is accompanied by several technical appendices. 
Appendix \ref{Supertwistors} includes essential definitions concerning the supergroup  $\sOSp(\mathcal{N}|4;\mathbb{R})$ and corresponding supertwistors. 
Appendix \ref{Stereographic} provides a review of the conformally flat atlas for AdS${}_d$.
In appendix \ref{AppendixC}, we spell out the $\cN$-extended superconformal algebra. 

Our two-component spinor notation and conventions follow 
\cite{BK}, and are similar to those adopted in  \cite{WB}.  The only difference is that the spinor Lorentz generators $(\s_{ab})_\a{}^\b$ and 
$({\tilde \s}_{ab})^\ad{}_\bd$  used in \cite{BK} have an extra minus sign as compared with \cite{WB}, 
specifically $\s_{ab} = -\frac{1}{4} (\s_a \tilde{\s}_b - \s_b \tilde{\s}_a)$ and 
 $\tilde{\s}_{ab} = -\frac{1}{4} (\tilde{\s}_a {\s}_b - \tilde{\s}_b {\s}_a)$.


\section{The (bi)supertwistor description of AdS$^{4|4\N}$} \label{section2}

In this section we give a brief review of the construction of \cite{KT-M21}. The reader is referred to appendix \ref{Supertwistors} for the technical details concerning the supergroup  $\sOSp(\mathcal{N}|4;\mathbb{R})$ and supertwistors.

Associated with the space of even complex supertwistors, ${\mathbb C}^{4|\cN}$, 
is a Grassmannian of even two-planes.
Given such a two-plane $\tilde \cP$, it is spanned by two even supertwistors 
$T^\m$,
\bea
\cP = \big( T^\m \big)= (T_A{}^\m) ~, \qquad \m = 1,2~.
\label{two-plane}
\eea
The property of $\cP$ being a two-plane means that the bosonic bodies of $T^1$ and $T^2$ are linearly independent complex four-vectors. 
 An arbitrary element $Q \in \tilde \cP$ is a linear combination $Q= T^\m q_\m$, with the coefficients $q_\m$ being even elements of the Grassmann algebra.  
By construction,  the supertwistors \eqref{two-plane} are defined modulo the equivalence relation
\begin{align} \label{equiv}
	T^\mu \sim T'^\m=T^\nu R_\nu{}^\mu\,, \qquad R=(R_\m{}^\n)  \in \sGL(2,\mathbb{C})\,,
\end{align}
since both $T^\m$ and $T'^\m $ define the same two-plane $\tilde \cP$.

We restrict our attention to the subset of those two-planes which satisfy the constraints 
\begin{subequations}\label{planecon}
\bea
\det \big( \cP^{\rm sT} {\mathbb J} \cP \big) &\neq &0~, \label{planecon.a}\\
\cP^\dagger {\mathbb J} \cP \equiv (* \cP)^{\rm sT} {\mathbb J} \cP &=& 0~.\label{planecon.b}
\eea
\end{subequations}
Here \eqref{planecon.a} refers to the body of the $2\times 2$ supermatrix 
$ \cP^{\rm sT} {\mathbb J} \cP $, and $*T$ denotes the conjugate of a pure supertwistor $T$, see eq. \eqref{A.14}. 
The conditions \eqref{planecon} imply that the bodies of the four even supertwistors 
$T^{\hat \m} = (T^\m, *T^{\dot \m} ) $
form a basis for ${\mathbb C}^4$, in particular the supertwistors \eqref{two-plane} generates a two-plane.
We emphasise that the conditions \eqref{planecon} are invariant under the equivalence transformations \eqref{equiv}.
In what follows, the supertwistor $*T$ will be denoted $\bar{T}$. 

We say that any pair of even supertwistors $\cP$, eq. \eqref{two-plane},
 constrained by the conditions \eqref{planecon} constitutes a frame.
The space of frames will be denoted ${\mathfrak F}_\cN$. The supergroup $ \sOSp(\cN|4; {\mathbb R})$ acts on ${\mathfrak F}_\cN$ by the rule 
\bea
g (T^\m , \bar T^{\dot \m} ) = (g T^\m , g  \bar T^{\dot \m} )~, 
\qquad g \in \sOSp(\cN|4; {\mathbb R})~.
\label{2.4}
\eea
This group action is naturally extended to the quotient space 
${\mathfrak F}_\cN / \sim$. The latter proves to be a homogeneous space of 
$ \sOSp(\cN|4; {\mathbb R})$, which was  identified in  \cite{KT-M21} with the AdS superspace, 
\bea
\rm{AdS}^{4|4\cN}= {\mathfrak F}_\cN/\sim~.
\label{3.11} 
\eea

Given two frames $T^{\hat \m} , \widetilde{T}^{\hat \m} \in  {\mathfrak F}_\cN$, 
one can construct the following
$\sOSp(\cN|4; {\mathbb R})$-invariant two-point functions:
\begin{subequations}\label{two-point}
\bea
\frac{1}{\ell^2}
\o( T, \widetilde T) &:=& -2\frac{ \langle \bar  T^{\dot \m} | \widetilde{T}^\n \rangle 
 \langle  \bar T_{\dot \m} | \widetilde{T}_\n \rangle   } 
{ 
\langle  \bar T^{\dot \s} | \bar T_{\dot \s} \rangle 
\langle  \widetilde{T}^\r | \widetilde{T}_\r \rangle
}~, \label{two-point.a}
\\
\frac{1}{\ell^2}
\o_{(+)} ( T, \widetilde T) &:=& 2\frac{ \langle  T^{ \m} | \widetilde{T}^\n \rangle  
 \langle  T_{ \m} | \widetilde{T}_\n \rangle   } 
{ 
\langle  T^{ \s} |  T_{ \s} \rangle
\langle  \widetilde{T}^\r | \widetilde{ T}_\r \rangle } -1~,  \label{two-point.b}\\
\frac{1}{\ell^2}
\o_{(-)}( T, \widetilde T) &:=& 2\frac{ \langle  \bar T^{\dot \m} | \widetilde{\bar T}{}^{\dot \n} \rangle
 \langle  \bar T_{\dot \m} | \widetilde{\bar T}_{\dot \n} \rangle   } 
{ 
\langle  \bar T^{\dot \s} | \bar T_{\dot \s} \rangle
\langle  \widetilde{\bar T}{}^{\dot \r} | \widetilde{\bar T}_{\dot \r} \rangle 
}-1 ~,  \label{two-point.c}
\eea
\end{subequations}
with $\ell$ a fixed positive parameter.
They do not change if $T$ and $\widetilde T$ are  replaced by their equivalent frames \eqref{equiv}, and therefore these $\sOSp(\cN|4; {\mathbb R})$-invariant two-point functions are well defined on $\rm{AdS}^{4|4\cN}$. In the non-supersymmetric case, 
$\cN=0$, the three two-point functions \eqref{two-point} coincide.

Given a point in  ${\mathfrak F}_\cN$, we associate with it the  graded antisymmetric matrices
\begin{subequations}\label{bi-super}
\bea
X_{AB} &:=& -2 \ell \frac{ T_A{}^{ \m}  {T}_{B}{}^\n \ve_{\m\n} } 
{ \langle   T^{ \g} |  T^{\d} \rangle \ve_{\g \d} }
=- (-1)^{\e_A \e_B}  X_{BA}~,
\\
\bar X_{AB} &:=& -2 \ell\frac{ \bar T_A{}^{ \dmu}  \bar T_B{}^{\dnu}  \ve_{\dmu \dnu}  } 
{ \langle   \bar T^{ \dot \g} |  \bar T^{\dot \d} \rangle  \ve_{\dot \g \dot \d}}
=- (-1)^{\e_A \e_B}  \bar X_{BA}~.
\eea
\end{subequations} 
These supermatrices are invariant under arbitrary equivalence transformations \eqref{equiv},
and therefore they may be used to parametrise AdS$^{4|4\cN} $.
The bi-supertwistors \eqref{bi-super} have the following properties: 
\begin{subequations} \label{bi-super2}
\bea
X_{[AB} X_{CD \}} &=&0~,  \label{bi-super2.a}\\
(-1)^{\e_B} X_{AB} {\mathbb J}^{BC} X_{CD} &=& \ell X_{AD} ~,\\
 {\mathbb J}^{BA}  X_{AB}&=& 2\ell~, \\
(-1)^{\e_B} X_{AB} {\mathbb J}^{BC} \bar X_{CD} &=& 0~,
\eea
\end{subequations}
where $X_{[AB} X_{CD \}} $ denotes the graded antisymmetric 
part of 
$X_{AB} X_{CD } $.
Using the results of  \cite{K-compactified12}, the bi-supertwistor description of AdS$^{4|4\cN}$ defined by \eqref{bi-super2} may be shown to be equivalent to the supertwistor one described earlier.

Restricting the above bi-supertwistor realisation of AdS$^{4|4\cN}$ to the $\cN=0$ case gives the bi-twistor formulation of AdS$_4$, which in turn 
leads to a standard embedding formalism for AdS$_4$. Building on the analysis given in section 3.3 of \cite{KT-M21}, it may be used to derive the reality condition
\bea
X_{\langle \hal \hbe \rangle} 
+ {\bar X}_{\langle \hal \hbe \rangle } =0 ~,\qquad 
X_{\langle \hal \hbe \rangle  } := X_{ \hal \hbe } - \frac{\ell}{2} J_{\hal \hbe}~, \quad 
\bar X_{\langle \hal \hbe \rangle  } := \bar X_{ \hal \hbe } - \frac{\ell}{2} J_{\hal \hbe}~.
\eea
Here $X_{\langle \hal \hbe \rangle} $ and $\bar X_{\langle \hal \hbe \rangle} $ denote 
the $J$-traceless parts of $X_{ \hal \hbe } $ and $\bar X_{ \hal \hbe   } $, respectively,
\bea
J^{ \hal \hbe } X_{\langle \hal \hbe \rangle} =0~, \qquad 
J^{ \hal \hbe } \bar X_{\langle \hal \hbe \rangle} =0~.
\eea
Associated with $X_{ \hal \hbe } $ is a {\it real} 5-vector 
\bea
X_{\hat a} := \frac{\ri }{2} (J\G_{\hat a} )^{\hal \hbe} X_{\langle \hal  \hbe \rangle } 
=\frac{\ri }{2} (J\G_{\hat a} )^{\hal \hbe} X_{ \hal \hbe } 
~, \qquad \hat a = 0,1,2,3,4 ~.
\eea
Here $\G_{\hat a}= \big( (\G_{\hat a})_\hal{}^\hbe\big)$ are {\it real} $4\times 4$ matrices which obey the anti-commutation relations 
\bea
\{ \G_{\hat a}  , \G_{\hat b}  \} = 2\eta_{\hat a \hat b} {\mathbbm 1}_4~, \qquad 
\eta_{\hat a \hat b}= {\rm diag} \, (-1,+1,+1,+1,-1)~, 
\eea
and are characterised by the property
\bea
\G_{\hat a}^{\rm T} = J  \G_{\hat a}  J^{-1} \quad \Longleftrightarrow \quad
(J  \G_{\hat a})^{\rm T} = -J  \G_{\hat a} ~. 
\eea
The explicit realisation of 
$\G_{\hat a}$ is given, e.g., in \cite{KPT-MvU}.
Making use of the completeness relation
\bea
(J\, \G^{\hat a})^{\hat \a \hat \b} (J\, \G_{\hat a} )^{\hat \g \hat \d} 
= - J^{\hat \a \hat \b} J^{\hat \g \hat \d} 
+ 2 (J^{\hat \a \hat \g} J^{\hat \b \hat \d} - J^{\hat \a \hat \d} J^{\hat \b \hat \g} )~,
\eea 
we obtain 
\bea
X^{\hat a} X_{\hat a} = - \ell^2~.
\eea
It may be shown that the bi-twistor description of AdS$_4$ is equivalent to the bi-spinor formalism introduced in \cite{BFP}.

Since the two-point functions \eqref{two-point}  are invariant under arbitrary equivalence transformations \eqref{equiv}, they can be expressed in terms of the bi-supertwistors
\eqref{bi-super}. In terms of the supermatrices $X = (X_A{}^B) $ and $\bar X= (\bar X_A{}^B)$ defined by 
\bea
X_A{}^B = (-1)^{\e_C} X_{AC} {\mathbb J}^{CB}~,\qquad 
\bar X_A{}^B = (-1)^{\e_C} \bar X_{AC} {\mathbb J}^{CB}~,
\eea 
these expressions have the form:
\begin{subequations} \label{str two-points}
\bea
\o( T, \widetilde T) &=& -\frac 12 {\rm Str} \big( \bar X \widetilde{X} \big)~, \\
\o_{(+)}( T, \widetilde T) &=& \frac 12 {\rm Str} \big( X \widetilde{X} \big)-\ell^2~,\\
\o_{(-)}( T, \widetilde T) &=&\frac 12 {\rm Str} \big( \bar X \widetilde{\bar X} \big)-\ell^2~.
\eea
\end{subequations}
We point out that the $ \sOSp(\cN|4; {\mathbb R})$ transformation \eqref{2.4} acts on $X$ and $\bar X$ as follows 
\bea
X~ \to ~ g Xg^{-1} ~, \qquad \bar X~ \to ~ g \bar Xg^{-1} ~.
\eea

The bi-supertwistor realisation described above facilitates the construction of manifestly $\sOSp(\N|4;\mathbb{R})$ invariant models. 
Indeed, let us consider the following worldline action for a superparticle on AdS$^{4|4\N}$ 
\begin{align} \label{superparticle model}
	S = - \hf \int \rd \t  \mathfrak{e}^{-1}\Bigg \{  \frac {1}{4} {\rm Str} \big(\dot{\bar{X}}\dot{X}\big) + \ri \k \left( {\rm Str}\big(\dot{X}\dot{X}\big) - {\rm Str}\big(\dot{\bar{X}}\dot{\bar{X}}\big)\right) + (\mathfrak{e} m)^2 \Bigg \}~,
\end{align}
where $\t$ parametrises the world line, $\mathfrak{e}$ denotes the einbein, $\k$ is a real dimensionless parameter, and $m$  is a mass parameter. We can see that in the non-supersymmetric case, $\N=0$, 
the $\k$-term is absent, 
since the three two-point functions \eqref{two-point} coincide.


\section{Isomorphic realisation of the AdS supergroup} \label{section3}

The supergroup $\sOSp(\mathcal{N}|4;\mathbb{R})$ possesses an alternative realisation, 
which we introduce below and which turns out to be useful for applications.
There is a simple motivation to look for such a realisation. To explain it, we consider the non-supersymmetric case, $\cN=0$. It follows from \eqref{planecon.a} 
that for every frame 
\begin{align} 
	{\cP}
	= \left( \begin{array}{c}
		F \\
		G 	
		\end{array} \right) \in {\mathfrak F}_0~, \qquad F, G \in {\sMat} (2, {\mathbb C}) ~,
\end{align}
the  $2\times 2$ matrices $F$ and $G$ are non-zero. In the framework of the coset construction, however, it would be useful to deal with an isomorphic realisation of $\sSp(4,\mathbb{R})$ that would allow a frame such that either $F=0$ or $G=0$ .


Let us consider a supergroup, denoted  $\sOSp(\mathcal{N}|4;\mathbb{R})_{C}$,
 consisting of 
all even $(4|\cN)\times(4|\cN)$ supermatrices $\underline{g}$ subject to the following constraints:
\bsubeq  \label{group constraints}
\begin{align}
	\underline{g}^{\sT} \mathbb{K} \underline{g} &=\mathbb{K}\,, \label{gkg}\\
	\underline{g}^\dag \mathbb{J} \underline{g} &= \mathbb{J} \label{gdagjg}\,.
\end{align}
\esubeq
Here we have introduced the graded antisymmetric $(4|\cN)\times (4|\cN)$ supermatrix
\begin{align} \label{kdef}
	\mathbb{K} = \left(
	\begin{array}{c||c}
		\hat{\ve} ~& ~0\\
		\hline \hline
		0 ~&~ {\mathbbm 1}_\mathcal{N}
	\end{array}
	\right)\,, 
	\qquad 
	\hat{\ve} = \left(
	\begin{array}{cc}
		\ve & ~~0\\
		0 &~ -\ve^{-1}
	\end{array}
	\right)\,,
\end{align}
with
\begin{align}
	\ve 
	= \left(\begin{array}{rc}
		0~ & 1 \\
		-1 ~& 0
	\end{array}\right)\,.
	\label{epsilons}
\end{align}
In what follows, we will denote the components  of the matrices \eqref{epsilons} as
$\ve = (\ve^{\a\b}) $ and $\ve^{-1} = (\ve_{\a\b}) $, which is why we prefer to use
the notation $\ve^{-1}$ instead of $-\ve$.

 The supergroup $\sOSp(\mathcal{N}|4;\mathbb{R})_{C}$ 
proves to be isomorphic to $\sOSp(\mathcal{N}|4;\mathbb{R})$. 
The proof is based on considering the following supermatrix correspondence:
\bsubeq
\begin{align}
	g &\rightarrow \underline{g} := \mathbb{U}g\mathbb{U}^{-1}\,, \qquad 
	\forall g \in \sOSp(\mathcal{N}|4;\mathbb{R})\,,
\end{align} 
in conjunction with the supertwistor transformation
\begin{align}
	T &\rightarrow \underline{T} := \mathbb{U}T\,, \label{transformed t}
\end{align}
\esubeq
for every supertwistor $T$. Here the supermatrix $\mathbb{U}$ is defined as
\begin{align}
	\mathbb{U} = \left(
	\begin{array}{c||c}
		t_s & 0\\
		\hline \hline
		0 & {\mathbbm 1}_\mathcal{N}
	\end{array}
	\right)\,, 
	\qquad 
	t_s = \frac{1}{\sqrt{2}} \left(
	\begin{array}{cc}
		{\mathbbm 1}_2 & -\ri\ve^{-1}\\
		-\ri\ve & {\mathbbm 1}_2
	\end{array}
	\right)\,~.
\end{align}
It obeys the useful properties:
\begin{align} \label{ujk properties 1}
	\bar{\mathbb{U}}  = \mathbb{U}^{-1} \,, \qquad
	\mathbb{U}^{\sT} = 
	(\mathbb{U}^{-1})^\dag\,,
\end{align}
and
\bsubeq \label{ujk properties 2}
\begin{align} 
	(\mathbb{U}^{-1})^{\sT} \mathbb{J}\mathbb{U}^{-1} &= \ri\mathbb{K}\,, \label{ujuk} \\
	\mathbb{U}^{\sT} \mathbb{J} \mathbb{U}^{-1} &= \mathbb{J}\,. \label{ujuj}
\end{align}
\esubeq
These conditions imply that 
\begin{align} \label{uunitary}
	\mathbb{U}^\dag \mathbb{J} \mathbb{U} = \mathbb{J}\,.
\end{align}
Associated with $\sOSp(\mathcal{N}|4;\mathbb{R})_{C}$ are two invariant inner products defined as
\bsubeq \label{innerproducts}
\begin{align}
	\braket{\underline{T}}{\underline{S}}_\mathbb{K} &:= \underline{T}^{\sT}\mathbb{K}\underline{S}\,, \label{tkt} \\ 
	\braket{\underline{T}}{\underline{S}}_\mathbb{J} &:= \underline{T}^\dag \mathbb{J} \underline{S}\,, \label{tjs}
\end{align}
\esubeq
for arbitrary pure supertwistors $\underline{T}$ and $\underline{S}$. 
The conditions \eqref{group constraints} impose restrictions on the blocks of $\underline{g}$. For
\begin{align}
	\underline{g} &= \left( \begin{array}{c||c}
		A & B \\
		\hline
		\hline
		C & D
	\end{array} \right)\,,
\end{align}
these are:
\bsubeq \label{groupreq1}
\begin{align}
	A^\T \hat{\ve} A -  C^\T C &= \hat{\ve}\,, \label{aka con}
	\\
	B^\T \hat{\ve} B +  D^\T D &= {\mathbbm 1}_N\,, \label{bkb con}
	\\
	A^\T \me B -  C^\T D &= 0\,, \label{akb con}
\end{align}
\esubeq
and
\bsubeq \label{groupreq2}
\begin{align}
	A^\dag J A + \ri C^\dag C = J\,,
	\\
	B^\dag J B + \ri D^\dag D = \ri{\mathbbm 1}_N\,,
	\\
	A^\dag J B + \ri C^\dag D = 0\,.
\end{align}
\esubeq
In the original realisation of $\sOSp(\mathcal{N}|4;\mathbb{R})$ the reality condition could be realised as the coincidence of the supertranspose and the Hermitian conjugate, eq. \eqref{A.13b}.
For our new realisation of the supergroup, 
 \eqref{A.13b} is replaced with the following condition
\begin{align}
	\underline{g}^{\dag} = (\mathbb{U}^{\text{sT}})^{2}\underline{g}^{\text{sT}}(\mathbb{U}^{\dag})^{2}\,.
\end{align}
From this we have the following conditions
\bsubeq
\begin{align}
	\bar{A} &= t_s{}^{-2} A t_s{}^2\,, \label{areal}
	\\
	\bar{B} &= t_s{}^{-2} B\,, \label{breal}
	\\
	\bar{C} &= -C t_s{}^2\,, \label{creal}
	\\
	\bar{D} &= D\,. \label{dreal}
\end{align}
\esubeq

We will now discuss involution for the supertwistors $\underline{T}$. Since the transformation \eqref{transformed t} applies to every supertwistor $T$, we can also consider it applied to $*T$. We have
\begin{align}
	*T \rightarrow \underline{*T} = \mathbb{U}(*T)\,. 
\end{align} 
This acts explicitly on a supertwistor $\underline{T}$ as
\begin{align} \label{davidstar}
	\underline{T} = \left(\begin{array}{c}
		f \\
		g \\
		\hline\hline
		\ri\psi
	\end{array} \right) ~\rightarrow~
\underline{*T} = -\ri  \left(\begin{array}{c}
	\ve^{-1}\bar{g} \\
	 \ve\bar{f} \\
	\hline\hline
	(-1)^{1+\e({\underline{T}})}\bar{\psi}
\end{array}\right)\,.
\end{align}
The components of $\underline{*T}$ are given by
\begin{align}
	\underline{*T}_{A} = (-1)^{\e({\underline{T}})\e_{A}+\e_{A}}\overline{(\mathbb{U}^{-2}\underline{T})_{A}}\,.
\end{align}

Let us introduce a new operation, denoted by $\star$, by removing the factor of $-\ri$ in \eqref{davidstar}:
\begin{align}
	\underline{T} = \left(\begin{array}{c}
		f \\
		g \\
		\hline\hline
		\ri\psi
	\end{array} \right) \rightarrow
	\star\underline{T} = \left(\begin{array}{c}
		\ve^{-1}\bar{g} \\
		\ve\bar{f} \\
		\hline \hline 
		(-1)^{1+\e({\underline{T}})}\bar{\psi}
		\end{array}\right)\,.
\end{align}
The components of $\star\underline{T}$ are given by
\begin{align}
	(\star\underline{T})_{A} = (-1)^{\e({\underline{T}})\e_{A}+\e_{A}}\ri\overline{(\mathbb{U}^{-2}\underline{T})_{A}}\,.
\end{align}
We therefore have the following reality condition with respect to the map $\star$
\begin{align}
	\overline{\underline{T}_A} = (-1)^{\e(\underline{T})\e_{A}+\e_{A}}(-\ri)(\mathbb{U}^{-2}\underline{T})_{A}\,.
\end{align}
The map $\star$ is an involution, since it satisfies the property
\begin{align}
	\star(\star\underline{T}) = \underline{T}\,. 
\end{align}
We also observe that 
\begin{align}
	(\star\underline{T})^{\sT} = \ri\underline{T}^{\dag}(\mathbb{U}^{\sT})^{2}\,,
\end{align}
which, in conjunction with the properties \eqref{ujk properties 2}, yields the following
\begin{align}
	\braket{\star\underline{T}}{\underline{S}}_\mathbb{K} = (\star\underline{T})^{\sT}\mathbb{K}\underline{S} = \underline{T}^{\dag}\mathbb{J}\underline{S} = \braket{\underline{T}}{\underline{S}}_\mathbb{J}\,.
\end{align}

It is useful to express the constraints \eqref{planecon}, the two-point functions \eqref{two-point}, and the bi-supertwistors \eqref{bi-super} in terms of the new realisation of the supergroup. 
The constraints can be expressed as
\bsubeq \label{newcon}
\begin{align}
	\ve_{\m\n} \braket{\underline{T}^{\m}}{\underline{T}^{\n}}_{\mathbb{K}} &\neq 0 \,,
	\\
	\braket{\underline{T}^{\m}}{\underline{T}^{\n}}_{\mathbb{J}} &= 0\,. 
\end{align}
\esubeq
For the two-point functions we find
\bsubeq \label{new two-point}
\begin{align} \label{ads two point function}
	\frac{1}{\ell^2} \o(\underline{T},\underline{\widetilde{T}}) &= -2 \frac{\braket{\underline{T}^\mu}{\underline{\widetilde{T}}{}^{\nu}}_\mathbb{J}\braket{\underline{T}_\mu}{\underline{\widetilde{T}}_{\nu}}_\mathbb{J}}{\braket{\star\underline{T}{}^{\dot{\s}}}{\star\underline{T}{}_{\dot{\s}}}_\mathbb{K}
		\braket{\underline{\widetilde{T}}{}^{\s}}{\underline{\widetilde{T}}{}_{\s}}_\mathbb{K}}\,,
	\\
	\frac{1}{\ell^2} \o_{(+)}(\underline{T},\underline{\widetilde{T}}) &= 2 \frac{\braket{\underline{T}^\mu}{\underline{\widetilde{T}}{}^{\nu}}_\mathbb{K}\braket{\underline{T}_\mu}{\underline{\widetilde{T}}_{\nu}}_\mathbb{K}}{\braket{\underline{T}{}^{\s}}{\underline{T}{}_{\s}}_\mathbb{K}
		\braket{\underline{\widetilde{T}}{}^{\r}}{\underline{\widetilde{T}}{}_{\r}}_\mathbb{K}}-1\,,
	\\
	\frac{1}{\ell^2} \o_{(-)}(\underline{T},\underline{\widetilde{T}}) &= 2 \frac{\braket{\star\underline{T}^{\dot{\mu}}}{\star\underline{\widetilde{T}}{}^{\dot{\nu}}}_\mathbb{K}\braket{\star\underline{T}_{\dot{\mu}}}{\star\underline{\widetilde{T}}_{{\dot{\nu}}}}_\mathbb{K}}{\braket{\star\underline{T}{}^{\dot{\s}}}{\star\underline{T}{}_{\dot{\s}}}_\mathbb{K}
		\braket{\star\underline{\widetilde{T}}{}^{{\dot{\r}}}}{\star\underline{\widetilde{T}}{}_{{\dot{\r}}}}_\mathbb{K}} -1\,.
\end{align}
\esubeq
The bi-supertwistors \eqref{bi-super} can be expressed in terms of transformed supertwistors $\underline{T}$ as follows
\bsubeq \label{new bi-super}
\begin{align}
	\underline{X} &= (\underline{X}_{AB}) = \mathbb{U}X\mathbb{U}^{\sT} \,,
	\\
	\underline{\bar{X}} &= (\underline{\bar{X}}_{AB}) = \mathbb{U}\bar{X}\mathbb{U}^{\sT}\,. 
\end{align}
\esubeq
They satisfy the following properties
\bsubeq
\begin{align}
	\underline{X}_{[AB}\underline{X}_{CD\}} &= 0\,,
	\\
	(-1)^{\e_{B}} \underline{X}_{AB}\mathbb{K}^{BC}\underline{X}_{CD} &= -\ri\ell\underline{X}_{AD}\,,
	\\
	\mathbb{K}^{BA}\underline{X}_{AB} &= -2\ri\ell\,,
	\\
	(-1)^{\e_{B}}\underline{X}_{AB}\mathbb{K}^{BC}\underline{\bar{X}}_{CD} &= 0\,. 
\end{align}
\esubeq
For the case $\N=0$, the $\me$-traceless parts of the bi-supertwistors take the form
\begin{align}
	\underline{X}_{\langle\hat{\a}\hat{\b}\rangle} = \underline{X}_{\hat{\a}\hat{\b}} + \frac{\ri}{2}\ell\me_{\hat{\a}\hat{\b}}\,, \qquad  \underline{\bar{X}}_{\langle\hat{\a}\hat{\b}\rangle} = \underline{\bar{X}}_{\hat{\a}\hat{\b}} + \frac{\ri}{2} \ell\me_{\hat{\a}\hat{\b}}\,.
\end{align}
As before, we can express the two-point functions \eqref{new two-point} in terms of the supermatrices $\underline{X}_{A}{}^{B}$ and $\underline{\bar{X}}_{A}{}^{B}$ defined by
\begin{align} \label{new raised Xs}
	\underline{X}_{A}{}^{B} = (-1)^{\e_{C}}\underline{X}_{AC}\mathbb{K}^{CB}\,, \qquad \underline{\bar{X}}_{A}{}^{B} = (-1)^{\e_{C}}\underline{\bar{X}}_{AC}\mathbb{K}^{CB}\,. 
\end{align}
They then take the form
\bsubeq \label{new trace functions}
\begin{align}
	\o(\underline{T},\underline{\widetilde{T}}) &= -\frac{1}{2}{\rm Str}\big(\underline{\bar{X}}\underline{\widetilde{X}}\big)\,,
	\\
	\o_{(+)}(\underline{T},\underline{\widetilde{T}}) &= \frac{1}{2}{\rm Str}\big(\underline{X} \underline{\widetilde{X}}\big) - \ell^{2}
	\,,
	\\
	\o_{(-)}(\underline{T},\underline{\widetilde{T}}) &= \frac{1}{2}{\rm Str}\big(\underline{\bar{X}} \underline{\widetilde{\bar{X}}}\big) - \ell^{2}\,.
\end{align}
\esubeq


%
\section{Coset construction} \label{section4}

The alternative realisation of the AdS supergroup described in the previous section is ideal for developing a coset construction for ${\rm AdS}^{4|4\cal N}$. 
To start with, it is worth recalling some basic definitions, see e.g. \cite{Kirillov} for more details. Consider a homogeneous space $X= G/H_{x_0}$, where $G$ is a Lie group and $H_{x_0}$ is the isotropy subgroup (or stabiliser) of some point $x_0 \in X$. A global coset representative is an injective map $\cS: X \to G$ such that $\p \circ \cS = {\rm id}_X$, where $\p$ denotes the natural (canonical)  projection 
$\p: G\to G/H_{x_0}$. For many homogeneous spaces, no global coset representative exists.
In such a case, local coset representatives $\cS_A :U_A \to G$ with the property $\p \circ \cS_A = {\rm id}_{U_A}$ can be introduced on open charts $\{ U_A\}$ that provide an atlas for $X$. In the intersection of two charts $U_A$ and  $U_B$, $U_A \cap U_B \neq \emptyset $,
the corresponding coset representatives  $\cS_A $ and  $\cS_B $ are related by a little group transformation, 
 $\cS_B (x) =   \cS_A (x) h_{AB} (x)$, with $h_{AB} (x) \in H_{x_0}$. 


\subsection{Isotropy subgroup}

As a marked (preferred) point $\underline{\cP}^{(0)}$
of AdS$^{4|4\mathcal{N}}$, we choose
\begin{align} \label{prefpoint}
	\underline{\cP}^{(0)} = \left( \begin{array}{c}
		{\mathbbm 1}_2\\
		0\\
		\hline \hline
		0
	\end{array} \right)\,.
\end{align}
The stabiliser $H$ of $\underline{\cP}^{(0)}$ consists of those elements $h$ of the AdS supergroup $\sOSp(\mathcal{N}|4;\mathbb{R})_{C}$
which satisfy the conditions
\begin{align}
	h \underline{\cP}^{(0)} = \left( \begin{array}{c}
		M  \\
		0\\
		\hline \hline
		0
	\end{array} \right)\,,
	\qquad M \in \sGL(2,\mathbb{C})\,.
\end{align}
These conditions imply that 
\begin{align}
	h = \left( \begin{array}{c||c}
		\begin{array}{cc}
			N & 0 \\
			0 & (N^\dagger)^{-1}
		\end{array}
		& 0 \\
		\hline \hline
		0 & R 
	\end{array} \right)\,, 
	\qquad N \in \sSL(2,\mathbb{C})\,, \quad R \in \sO(\mathcal{N})\,.
\end{align}
Thus the stability subgroup $H$ is isomorphic to 
\begin{align} \label{stabgroup}
	\sSL(2,\mathbb{C}) \times \sO(\mathcal{N})\,.
\end{align}
The bi-supertwistors \eqref{new bi-super} corresponding to the preferred point $\underline{\cP}^{(0)}$ take the form
\begin{align} \label{pref bi-super}
	\underline{X}^{(0)} = \left(\begin{array}{c|c||c}
		-\ri\ve^{-1} & 0 & 0
		\\
		\hline
		0 & ~0~ & ~0~ 
		\\ 
		\hline \hline
		0 & 0 & 0
	\end{array}\right) \,, \qquad \underline{\bar{X}}^{(0)} = \left(\begin{array}{c|c||c}
	0 & 0 & 0
	\\
	\hline 
	~0~ & \ri\ve & ~0~ 
	\\
	\hline \hline 
	0 & 0 & 0
\end{array}\right) \,.
\end{align}

\subsection{Generalised coset representative}

The freedom to perform arbitrary equivalence transformations \eqref{equiv}
can be used to fine-tune the conditions \eqref{newcon} to
\bsubeq \label{normalised conditions}
\begin{align} 
	\braket{\underline{T}^\mu}{\underline{T}^\nu}_\mathbb{K} &= \ell\ve^{\mu\nu}\,,
\qquad \ell = \bar \ell >0~,  \label{normalised conditions.a}	\\
	\braket{\underline{T}^\mu}{\underline{T}^\nu}_\mathbb{J} &= 0\,,
	 \label{normalised conditions.b}
\end{align}
\esubeq
for a fixed positive parameter $\ell$. Such a frame is said to be normalised. 
Under the condition \eqref{normalised conditions.a}, the equivalence relation 
 \eqref{equiv} turns into 
\begin{align} \label{equivalence relation}
	\underline{T}^\m \sim \underline{T}^\n N_\n{}^\m
	\,, \qquad N \in \sSL(2,\mathbb{C})\,.
\end{align}
The space of normalised frames will be denoted ${\mathfrak F}_\cN^{(\ell)}$.
Along with the definition \eqref{3.11} given earlier,
the $\cN$-extended AdS superspace can equivalently be defined as 
$\text{AdS}^{4|4\mathcal{N}} = \frak{F}_\mathcal{N}^{(\ell)}/ \!\! \sim$, where the equivalence relation is given by \eqref{equivalence relation}.

The conditions \eqref{normalised conditions} can be recast in terms of the two-plane
\begin{align} \label{two plane def}
	\underline{\cP} = ( \underline{T}^\mu )
	= \left( \begin{array}{c}
		F \\
		G \\
		\hline \hline
		\ri\Q
	\end{array} \right) 
\end{align}
and imply the following constraints:
\bsubeq \label{detcon}
\begin{align}
	\det F &+ \det G = \ell + \frac{1}{2}\text{tr}\left(\Q\ve^{-1} \Q^\T  \right)\,,  \label{detcon1}
	\\
	F^\dag G &- G^\dag F + \ri\Q^\dag \Q = 0\, . \label{detcon2}
\end{align}
\esubeq
Relation \eqref{detcon1} tells us that at least one of the $2\times 2$ matrices $F$ and $G$ is nonsingular.

Associated with the normalised two-plane  $\underline{\cP} $ is the following group element
\bsubeq \label{cosetrep}
\begin{align}
	S(\underline{\cP}) 
&= \left( \begin{array}{c||c}
		A & B \\
		\hline
		\hline
		C & D
	\end{array} \right)\,,	\\
	A &= \left( \begin{array}{cc} \label{ablock}
		F ~& \ve^{-1} \bar{G} \ve^{-1} \\
		G ~& \ve \bar{F} \ve^{-1} 
	\end{array}
	\right) 
	\,,
	\\
	C &= \Big(\begin{array}{cc}
		\ri\Q &~ -\bar{\Q}\ve^{-1}
	\end{array} \Big) 
	\,,
	\\
	D &= \Big( {\mathbbm 1}_\mathcal{N} -  C [A^\T \me A]^{-1}C^\T\Big)^{-\frac{1}{2}}\,,
	\label{dblockdef} \\
	B&= \me^{-1}(A^{-1})^\T C^\T D \label{bblockdef}
	\,.
\end{align}
\esubeq
The fundamental property of $S(\underline{\cP}) $ is that $ S(\underline{\cP}) \underline{\cP}^{(0)} = \underline{\cP}$, for any normalised two-plane 
$\underline{\cP} \in {\mathfrak F}_\cN^{(\ell)}$.
We point out that $D$ is symmetric, $D=D^\T$. The functional forms of the matrices $A$ and $C$ are fixed through the condition  $ S(\underline{\cP}) \underline{\cP}^{(0)} = \underline{\cP}$ and 
the reality conditions \eqref{areal} and \eqref{creal}.
 The remaining blocks are then fixed by the group requirements \eqref{groupreq1} and \eqref{groupreq2}.
It is possible to obtain alternate expressions for the blocks $D$ and $B$, which may be more suited to performing calculations. They take the following form 
\bsubeq
\begin{align}
	D &= \Big({\mathbbm 1}_{\N} + C\me^{-1}C^{\T}\Big)^{\frac{1}{2}} \,,
	\\
	B &= A\me^{-1}C^{\T}D^{-1} \,.
\end{align}
\esubeq
These expressions can be seen to coincide with \eqref{dblockdef} and \eqref{bblockdef} by using the group requirements and the general form for the inverse of a supermatrix. 

The group element $S(\underline{\cP})$ is characterised by the property 
\begin{align} \label{rep equiv}
	S(\underline{\cP}N) = S(\underline{\cP})\mathfrak{N}\,, \quad 
	\mathfrak{N} = \left(\begin{array}{cc||c}
		N & 0 & 0 \\
		0 & (N^\dagger)^{-1} & 0 \\
		\hline \hline 
		0 & 0 & {\mathbbm 1}_{\N}
	\end{array} \right)\,,
\end{align}
with $N \in \sSL(2,\mathbb{C})$. This relation means that $S(\underline{\cP})$ is not a genuine coset representative that is used in 
the coset construction.
However,  $S(\underline{\cP})$ will allow us to obtain a coset representative if we pick a single two-plane in each equivalence class.
This may be readily done in coordinate charts for $\text{AdS}^{4|4\mathcal{N}}$.
%


\subsection{AdS space ($\cN=0$)}

As noted above,
at least one of the $2\times 2$ matrices $F$ and $G$, see eq. \eqref{two plane def}, is nonsingular.
Therefore we can naturally introduce two coordinate charts for ${\mathfrak F}_\cN^{(\ell)}$
that provide an atlas. We define the north chart to consist of all normalised two-planes with $\det F \neq 0$. Similarly, the south chart is defined to consist of all normalised two-planes with $\det G \neq 0$.

In the north chart, we can use the freedom \eqref{equivalence relation}
to choose
$F \propto {\mathbbm 1}_2$, and then 
\begin{align} \label{north chart general}
	\underline{\cP} 
	= \l\left( \begin{array}{c}
		\ell {\mathbbm 1}_2 \\
		- \tilde{\bm x}
	\end{array}\right)
\,, \qquad 
\tilde{\bm x} = x^{m}\tilde{\s}_{m}\,, \qquad \tilde{\s}_m = ({\mathbbm 1}_2, - \vec \s )~,
\end{align}
where
$\l \neq 0$ is a parameter, and $\vec \s$ are the Pauli matrices. The constraints \eqref{detcon1} and \eqref{detcon2} give, respectively, 
\begin{subequations}
\bea 
\l^2 ( \ell^2 + \det \tilde{\bm x} ) &=& \ell ~~\implies ~~ \bar \l^2 ( \ell^2 + \det \tilde{\bm x}^\dagger ) = \ell~, \label{4.12a} \\
\tilde{\bm x}^\dagger &=& \tilde{\bm x}~.
\eea
\end{subequations}
It follows that $x^m$ is real and $x^2 := \eta_{mn} x^m x^n \neq \ell^2$. We also observe that $\l= \bar \l$ . Since there is still a remnant of the equivalence relation 
\eqref{equivalence relation}, 
$	\underline{T}^\m \sim - \underline{T}^\m$, 
it can be used to fix $\l > 0$. Then we observe that the coordinate chart 
is specified by 
\bea
x^2< \ell^2~, 
\eea
and the parameter $\l$ is given by 
\begin{align}
\l = \sqrt{ \frac{\ell}{\ell^{2}-x^2} }\,.
\end{align}
The real coordinates $x^{m}$ parametrise AdS$_{4}$ in the north chart. 
Direct calculation of the two-point function \eqref{ads two point function} in this chart yields
\begin{align} \label{bosonic interval}
	\o(\underline{T},\underline{T}+\text{d}\underline{T}) =  \frac{\ell^{4} \eta_{mn}\text{d}x^{m}\text{d}x^{n}}{(\ell^{2}-x^{2})^{2}}\,. 
\end{align}

In the south chart,  the gauge freedom \eqref{equivalence relation} can be used
to choose
$G \propto {\mathbbm 1}_2$, 
and then 
\begin{align} \label{south chart general}
	\underline{\cP} 
	= \g \left( \begin{array}{c}
		{\bm y} \\
		\ell {\mathbbm 1}_2
	\end{array}\right)
\,, \qquad 
{\bm y} = y^{m}{\s}_{m}\,, \qquad \s_m = ({\mathbbm 1}_2, \vec{\s})~,
\end{align}
for some parameter $\g \neq 0$. Now, repeating the north-chart analysis 
tells us that the local coordinates $y^m$ are real, and the following relations hold:
\bea
y^2 <\ell^2~, \qquad \g = \sqrt{ \frac{\ell}{\ell^{2}-y^2} }\,.
\eea
The two-point function \eqref{ads two point function} in the south chart is
\begin{align}
	\o(\underline{T},\underline{T}+\text{d}\underline{T}) =  \frac{\ell^{4} \eta_{mn}\text{d}y^{m}\text{d}y^{n}}{(\ell^{2}-y^{2})^{2}}\,. 
\end{align}

In the intersection of the two charts, the transition functions are
\bea
{\bm y} = - \ell^2  \tilde{\bm x}^{-1} ~~ \Longleftrightarrow ~~y^m =  \frac{\ell^2}{x^2} x^m~~\implies ~ ~y^2 x^2 = \ell^4~.
\label{4.19}
\eea
It follows that $x^2 <0 \Longleftrightarrow y^2 <0$.
Comparing the above relations with those described in appendix \ref{Stereographic},
we find complete agreement except for the sign difference \eqref{4.19} and \eqref{B.9}.


%
%
\subsection{$\N\neq0$} \label{n s cosets}
The analysis of the previous subsection can be extended to the supersymmetric case in a similar fashion. 
Let us consider the north chart in which the matrix $F$ in \eqref{two plane def}
is nonsingular. The 
equivalence relation \eqref{equivalence relation}
can once again be used to 
choose $F \propto {\mathbbm 1}_2$, and then
\begin{align} \label{chiral chart}
	\underline{\cP} 
	=\l (x_+, \q) \left(\begin{array}{c}
		\ell {\mathbbm 1}_2 \\
		-\tilde{\bm x}_{+} \\
		\hline \hline 
		2\ri \sqrt{\ell} \q
	\end{array}\right)
\,,\qquad
	 \tilde{\bm x}_+ = x^{m}_+\tilde{\s}_{m}\,.
\end{align}
Making use of  \eqref{normalised conditions} leads to the relations
\bsubeq \label{chiral chart2}
\begin{align}
	&\tilde{\bm x}_{+} - \tilde{\bm x}_{-} = 4\ri\q^{\dag}\q\,, 
	\qquad \tilde{\bm x}_{-} := (\tilde{\bm x}_{+})^{\dag}\,, 	\\
	&\l = \sqrt{ \frac{\ell}{\ell^{2} - x^2_{+} - 2\ell\q^2}}\,, \qquad \q^{2} := 
	\text{tr}\left(\q\ve^{-1} \q^\T  \right)
	\,. 
\end{align}
\esubeq
The former is solved by 
\begin{align}
	\tilde{\bm x}_{\pm} = x^{m}\Tilde{\s}_{m} \pm 2\ri\q^{\dag}\q\,.
\end{align}
We see that the two-planes \eqref{chiral chart} are parametrised by the chiral coordinates $x_{+}^{m}$ and $\q_{I}{}^{\m}$, with $x_{+}^{m} = x^{m} + \ri\q_I \s^{m}\bar{\q}_I$. 

The coset representative in the north chart is given by
\bsubeq
\begin{align}
	S(\underline{\cP})_{\text{north}} &= 
	\left(
	\begin{array}{c||c}
		A_{\text{n}} & B_{\text{n}} \\
		\hline\hline
		C_{\text{n}} & D_{\text{n}}
	\end{array}
	\right) \label{north chart coset rep}\,, 
	\\
	A_{\text{n}} &= 
	\left(
	\begin{array}{c|c}
		\l {\mathbbm 1}_{2} & -\bar{\l}\ve\tilde{x}_{-}^{\T}\ve \\ 
		\hline
		-\l \tilde{x}_{+} &  \bar{\l} {\mathbbm 1}_{2}
	\end{array}
	\right)
	\,,
	\\
	C_{\text{n}} &= 
	\left(
	\begin{array}{c|c}
		2\ri\l\q & ~2\bar{\l}\bar{\q}\ve
	\end{array}
	\right)
	\,,
	\\
	D_{\text{n}} &= \Big({\mathbbm 1}_{\N} + C_{\text{n}}\me^{-1}{C_{\text{n}}}^{\T}\Big)^{\frac{1}{2}} \,,
	\\
	B_{\text{n}} &= A_{\text{n}}\me^{-1}{C_{\text{n}}}^{\T}{D_{\text{n}}}^{-1}\,.
\end{align}
\esubeq
The two-point function \eqref{ads two point function} computed in the north chart yields 
\begin{align} \label{susy chiral interval}
	\o(\underline{T},\underline{T}+\text{d}\underline{T}) =  \frac{\ell^{4}\eta_{mn}e^{m}e^{n}}{(\ell^{2}-x_{+}^{2}-2\ell\q^{2})(\ell^{2}-x_{-}^{2}-2\ell\bar{\q}^{2})}\,,
\end{align}
where
\begin{align} 
	e^{m} = \text{d}x^{m} + \ri(\q\s^{m}\text{d}\bar{\q}
	- \text{d}\q\s^{m}\bar{\q})\,, \qquad x_{\pm}^{2} = \eta_{mn}x_{\pm}^{m}x_{\pm}^{n} \,.
	\label{4.27}
\end{align}
In the non-supersymmetric case, $\N=0$, this reduces to
\eqref{bosonic interval}.

In the south chart, the gauge freedom \eqref{equivalence relation} can be used to fix $G \propto {\mathbbm 1}_2$. Repeating the analysis of the north chart leads to
\begin{align} \label{south chart two planes}
	\underline{\cP} = \g (y_+, \x) 
	\left(
	\begin{array}{c}
		{\bm y}_{+} \\
		\ell {\mathbbm 1}_{2} \\
		\hline \hline
		2\ri\sqrt{\ell}\xi
	\end{array}
	\right) 
	\,,\qquad 
	{\bm y}_+ = y^{m}_+{\s}_{m}\,,
\end{align}
with
\bsubeq
\begin{align}
	&{\bm y}_{+} - {\bm y}_{-} = 4\ri\xi^{\dag}\xi\,, 
	\qquad {\bm y}_{-} := ({\bm y}_{+})^{\dag}\,,	\\
	&\g = \sqrt{ \frac{\ell}{\ell^{2} - y_{+}^{2} - 2\ell\xi^2}}\,, \qquad \xi^{2} := \text{tr}\left(\xi\ve^{-1} \xi^\T  \right)\,. 
\end{align}
\esubeq
The former is solved by 
\begin{align}
	{\bm{y}_{\pm}} = y^{m}\s_{m} \pm 2\ri\xi^{\dag}\xi\,.
\end{align}
We see that the two-planes in the south chart \eqref{south chart two planes} are parametrised by the chiral coordinates $y_{+}^{m}$ and $\xi_{I}{}^{\m}$, with $y_{+}^{m} = y^{m} + \ri\xi_I \tilde{\s}^{m}\bar{\xi}_I$. 

The coset representative in the south chart is given by
\bsubeq
\begin{align}
	S(\underline{\cP})_{\text{south}} &= 
	\left(
	\begin{array}{c||c}
		A_{\text{s}} & B_{\text{s}} \\
		\hline\hline
		C_{\text{s}} & D_{\text{s}}
	\end{array}
	\right) \label{south chart coset rep}\,, 
	\\
	A_{\text{s}} &= 
	\left(
	\begin{array}{c|c}
		\g y_{+} & -\bar{\g}{\mathbbm 1}_{2} \\ 
		\hline
		\g{\mathbbm 1}_{2} & - \bar{\g} \ve y_{-}^{\T} \ve
	\end{array}
	\right)
	\,,
	\\
	C_{\text{s}} &= 
	\left(
	\begin{array}{c|c}
		2\ri\g\xi & ~2\bar{\g}\bar{\xi}\ve
	\end{array}
	\right)
	\,,
	\\
	D_{\text{s}} &= \Big({\mathbbm 1}_{\N} + C_{\text{s}}\me^{-1}{C_{\text{s}}}^{\T}\Big)^{\frac{1}{2}} \,,
	\\
	B_{\text{s}} &= A_{\text{s}}\me^{-1}{C_{\text{s}}}^{\T}{D_{\text{s}}}^{-1}\,.
\end{align}
\esubeq
The two-point function \eqref{ads two point function} computed in the south chart yields 
\begin{align} \label{susy south interval}
	\o(\underline{T},\underline{T}+\text{d}\underline{T}) =  \frac{\ell^{4} \eta_{mn}e'^{m}e'^{n}}{(\ell^{2}-y_{+}^{2}-2\ell\xi^{2})(\ell^{2}-y_{-}^{2}-2\ell\bar{\xi}^{2})}\,,
\end{align}
where
\begin{align} 
	e'^{m} = \text{d}y^{m} + \ri(\xi\s^{m}\text{d}\bar{\xi}
	- \text{d}\xi\s^{m}\bar{\xi})\,, 
	\qquad y_{\pm}^{2} = \eta_{mn}y_{\pm}^{m}y_{\pm}^{n}~.
\end{align}
In the intersection of the two charts, the transition functions are given by
\bsubeq \label{susy transition functions}
\begin{align}
	{\bm y}_{+} &= -\ell^{2}\tilde{\bm x}_{+}^{-1} ~~ \Longleftrightarrow ~~ y_{+}^{m} = \frac{\ell^{2}}{x_{+}^{2}}x_{+}^{m} ~~ \implies ~ y_{+}^{2}x_{+}^{2} = \ell^{4}\,,
	\\
	\xi &= -\q\tilde{\bm x}_{+}^{-1} \,.
\end{align}
\esubeq
In addition, the two coset representatives \eqref{north chart coset rep} and \eqref{south chart coset rep} are related in the intersection by the point-dependent little group transformation
\begin{align} \label{s coset to n coset}
	S(\underline{\cP})_{\text{north}} &= S(\underline{\cP})_{\text{south}} h^{-1}(x_{+},\q) \,, \qquad h^{-1} \in H\,. 
\end{align}
Explicitly, $h^{-1}$ is given by
\bsubeq
\begin{align}
	h^{-1} &= \left(\begin{array}{c|c||c} \label{coset reps trf}
		n^{-1} & 0 & 0 \\
		\hline 
		0 & n^{\dag} & 0 \\
		\hline \hline
		0 & 0 & {\mathbbm 1}_{\N}
	\end{array}\right)\,, 
\\
n^{-1} &= -\l\g^{-1}\tilde{\bm x}_{+}\,. \label{n inverse block} 
\end{align}
\esubeq
We see that $n^{-1}$ is chiral, through the transition functions \eqref{susy transition functions}.

So far we have only considered the form of the two-planes in the north and south charts. It is also useful to describe the form of the bi-supertwistors \eqref{new bi-super} in an explicit coordinate system. In the north chart they take the form
\begin{align}
	\underline{X}_{AB} &= -\ri \l^{2} (x_+, \q) 
	\left(\begin{array}{c|c||c}
		\ell^{2}\ve_{\a\b} & - \ell {x}_{+ \a}{}^{\bd} & 2\ri\ell^{\frac{3}{2}}\q_{\a J}
		\\
		\hline 
		\ell {{x}}_{+}{}^{\ad}{}_{\b} & \ve^{\ad\bd}x_{+}^{2} & -2\ri\ell^{\frac{1}{2}}{{x}}_{+}{}^{\ad\m}\q_{\m J}
		\\
		\hline \hline
		-2\ri\ell^{\frac{3}{2}}\q_{I\b} & 2\ri\ell^{\frac{1}{2}}\q_{I\m}{{x}}_{+}{}^{\bd\m} & -4\ell\q_{I}{}^{\m}\q_{\m J}
	\end{array}\right)\,.
\end{align}
It is of interest to compare this supermatrix with a similar result for compactified 
$\cN$-extended Minkowski superspace, see eq. (3.17) in \cite{K-compactified12}.


\section{Superspace geometry} \label{section5} 

In this section we give explicit expressions for the vierbein, connection, torsion tensor and curvature tensor. From these expressions the graded commutation relations of the covariant derivatives can be derived. 


\subsection{Geometric structures in AdS$^{4|4\mathcal{N}}$} \label{geometric objects}
Let us denote by $\mathcal{G}$ the superalgebra of the AdS supergroup $\sOSp(\mathcal{N}|4;\mathbb{R})_{C}$, and by $\mathcal{H}$ the algebra of the stability group \eqref{stabgroup}. 
Let $\mathcal{W}$ be the complement of $\mathcal{H}$ in $\mathcal{G}$, $\mathcal{G}=\mathcal{H}\oplus\mathcal{W}$. 
The superalgebra $\mathcal{G}$ consists of even supermatrices 
\begin{align}
	\underline{\Omega} = 
	\left(\begin{array}{c|c||c}
		u_{\a}{}^{\b} & v_{\a\bd}&\ri \psi_{\a J}\\
		\hline
		-{v}^{\ad\b} \phantom{\Big|}& -(u^{\dag})^{\ad}{}_{\bd} & -\bar{\psi}^{\ad}{}_{J}\\
		\hline \hline 
		\ri\psi_{I}{}^{\b} & \bar{\psi}_{I\bd} & \Lambda_{IJ} 
	\end{array}\right)
			 \in \mathcal{G}\,,\quad u \in \mathfrak{sl} (2,{\mathbb C}) \,, 
			 \quad \L \in \mathfrak{so} (\cN)~,
\end{align}
		with 
		$v=v^{\dag}$.
		The elements $\underline{h} \in \mathcal{H}$ take the form
\begin{align} \label{stabform}
			\underline{h} = \left(\begin{array}{c|c||c}
				n_{\a}{}^{\b} & 0 &0 \\
				\hline
				0 & -(n^{\dag})^{\ad}{}_{\bd} & 0\\
				\hline \hline 
				0 & 0 & r_{IJ} 
			\end{array}\right)\,, \quad n \in \mathfrak{sl}(2,\mathbb{C})\,, \quad r \in \mathfrak{so}(\mathcal{N})\,.
\end{align}
		Additionally, the elements $\underline{w} \in \mathcal{W}$ take the form
\begin{align} \label{compel}
			\underline{w} = \left(\begin{array}{c|c||c}
				0 & v_{\a\bd}&\ri \psi_{\a J}\\
				\hline
				-{v}^{\ad\b} &0 & -\bar{\psi}^{\ad}{}_{J}\\
				\hline \hline 
				\ri\psi_{I}{}^{\b} & \bar{\psi}_{I\bd} &0
			\end{array}\right) \,.
\end{align}
		With the following row-vector definition
\begin{align}
			\varphi_{I}{}^{\hat{\a}} := \left(\begin{array}{c|c}
				\ri\psi_{I}{}^{\a} & \bar{\psi}_{I \ad}
			\end{array}\right)
\end{align}
		the elements \eqref{compel} take the form
\begin{align} \label{compform}
			\underline{w} = \left(\begin{array}{c||c}
				\begin{array}{c|c}
					0 & v_{\a\bd}  \\
					\hline
					-{v}^{\ad\b} & 0
				\end{array}&  \varphi_{\hat{\a} J} \\
				\hline \hline
				\varphi_{I}{}^{\hat{\b}} & 0 
			\end{array} \right) \,.
\end{align}
		It is straightforward to verify $[\mathcal{W},\mathcal{H}] \subset \mathcal{W}$. 
		
		We may uniquely decompose the Maurer-Cartan one-form $\omega = S^{-1}\text{d}S$ as a sum $\omega = \textbf{E} + \bf{\O}$, where $\textbf{E} = S^{-1}\text{d}S|_{\mathcal{W}}$ is the vierbein taking its values in $\mathcal{W}$. 
		The Maurer-Cartan one-form is
\begin{align} \label{mc1f}
			\omega = \left(\begin{array}{c||c}
				\o_{\text{Sp}(4)}  & \me^{-1} \cE^{\T} \\
				\hline \hline 
				\cE & \O_{\text{SO}(\N)}
			\end{array}\right)\,,
\end{align}
		where the blocks are given by
\bsubeq
\begin{align}
			\o_{\text{Sp}(4)} &= \left(\begin{array}{c|c}
				\ve^{-1} F^{\T}\ve\text{d}F - \ve^{-1} G^{\T}\ve^{-1} \text{d}G  & \ve^{-1} F^{\T}\text{d}\bar{G}\ve^{-1}  - \ve^{-1} G^{\T}\text{d}\bar{F}\ve^{-1}  \\
				+ \ve^{-1} \Q^{\T}\text{d}\Q
				& + \ri\ve^{-1} \Q^{\T}\text{d}\bar{\Q}\ve^{-1} \\
				\hline
				F^{\dag}\text{d}G - G^{\dag}\text{d}F &  F^{\dag}\ve\text{d}\bar{F}\ve^{-1}  - G^{\dag}\ve^{-1} \text{d}\bar{G}\ve^{-1} 
				\\
				+ \ri \Q^{\dag}\text{d}\Q & - \Q^{\dag}\text{d}\bar{\Q}\ve^{-1} 
			\end{array}\right)\,,
			\\
			\cE &= D\text{d}C - DC(A^{-1})\text{d}A\,, \label{vierbeinoddblock}
			\\
			\O_{\text{SO}(\N)} &= D^{-1}\text{d}D 
			-  DC[(A^{-1})\me^{-1}\text{d}(A^{-1})^{\T}C^{\T}
			+ (A^{-1})\me^{-1}(A^{-1})^{\T}\text{d}C^{\T}]D\,.
\end{align}
\esubeq
		One can make use of the group conditions \eqref{groupreq1} to recast $\cE$ in an equivalent form
		\begin{align} \label{new spinor vierbein}
			\cE &= D\text{d}C - D^{-1}C\me^{-1}A^{\T}\me\text{d}A\,,
			\\ 
			\notag 
			&= \left(\begin{array}{c|c}
				2 \ri (\textbf{E}_{\Q})_{I}{}^{\a} & 2(\bar{\textbf{E}}_{\bar{\Q}})_{I\ad}
			\end{array}\right) \,. 
		\end{align}
		In the above, $\textbf{E}_{\Q}$ is given by
		\begin{align} \label{e theta block general}
			\textbf{E}_{\Q} &= \frac{1}{2}D\text{d}\Q - \frac{1}{2} D^{-1}\Q\ve^{-1}(F^{\T}\ve\text{d}F - G^{\T}\ve^{-1}\text{d}G)
			\\
			\notag
			& \quad - \frac{1}{2} \ri D^{-1}\bar{\Q}\ve^{-1}(F^{\dag}\text{d}G-G^{\dag}\text{d}F)\,. 
		\end{align}

		The Maurer-Cartan one-form \eqref{mc1f} can be decomposed into supermatrices of the form \eqref{stabform} and \eqref{compform} to obtain the vierbein and connection. 
		The connection is 
\begin{align}
			\bf{\O} = 
			\left(\begin{array}{c||c}
				\hat{\O}_{\hat{\a}}{}^{\hat{\b}} &  0 \\ 
				\hline \hline 
				0 & \O_{\text{SO}(\mathcal{N})}{}_{IJ}
			\end{array}\right)
		=
		\left(\begin{array}{c|c||c}
			\O_{\a}{}^{\b} & 0 &0 \\
			\hline
			0 & -\bar{\O}^{\ad}{}_{\bd} & 0\\
			\hline \hline 
			0 & 0 & \O_{\text{SO}(\mathcal{N})}{}_{IJ}
		\end{array}\right)\,,
\end{align}
		where 
\bsubeq
\begin{align}
			\O &= \ve^{-1} F^{\T}\ve\text{d}F - \ve^{-1} G^{\T}\ve^{-1} \text{d}G + \ve^{-1} \Q^{\T}\text{d}\Q\,.
\end{align}
\esubeq
		It is possible that these expressions may be simplified by using an explicit form for $A^{-1}$, with $A$ given by \eqref{ablock}, however the above expressions appear most convenient for proving the required properties
\begin{align} \label{connection conditions}
			\text{tr}\,\O = 0\,, \qquad 
			\Omega_{\text{SO}(\mathcal{N})}^{\T} = -\Omega_{\text{SO}(\mathcal{N})}\,. 
\end{align}
The vierbein is 
\begin{align} \label{vierbein}
			\textbf{E} = \left(\begin{array}{c||c}
				\hat{\textbf{E}}_{\hat{\a}}{}^{\hat{\b}} &  \cE_{\hat{\a} J} \\ 
				\hline \hline 
				\cE_{I}{}^{\hat{\b}} & 0
			\end{array}\right)
		= \left( \begin{array}{c|c||c}
			0 & \textbf{E}_{\a\bd} & 2\ri(\textbf{E}_{\Q})_{\a J} \\
			\hline
			- {\textbf{E}}^{\ad\b} & 0 & -2(\bar{\textbf{E}}_{\bar{\Q}})^{\ad}{}_{J} \\
			\hline \hline 
			2\ri (\textbf{E}_{\Q})_{I}{}^{\b} & 2(\bar{\textbf{E}}_{\bar{\Q}})_{I\bd} & 0
		\end{array} \right)
		 \,,
\end{align}
		where 
\begin{align} \label{vierbeinblock}
			\tilde{\textbf{E}} &= (\textbf{E}^{\ad \b})=-F^{\dag}\text{d}G + G^{\dag}\text{d}F - \ri\Q^{\dag}\text{d}\Q \,,
\end{align}
		and $\cE$ is defined as in \eqref{vierbeinoddblock} or \eqref{new spinor vierbein}. 
		It is straightforward to show that \eqref{vierbeinblock} is Hermitian, using \eqref{detcon2}. 

		Using the above expressions we can now compute the torsion $\mathcal{T}$ and curvature $\mathcal{R}$. In accordance with the coset construction, they are defined as follows:
\begin{align} \label{torscurvdef}
			\mathcal{T} = \text{d}\textbf{E} - \bf{\O} \wedge \textbf{E} - \textbf{E}\wedge\bf{\O}\,, \qquad \mathcal{R} = \text{d}\bf{\O} - \bf{\O} \wedge \bf{\O}\,. 
\end{align}
		There exists another simple expression for both $\mathcal{T}$ and $\mathcal{R}$, given by
		\begin{align} \label{ewedgee}
			\mathcal{T} = (\textbf{E}\wedge\textbf{E})|_{\mathcal{W}}\,, \qquad \mathcal{R} = (\textbf{E}\wedge\textbf{E})|_\mathcal{H}\,. 
		\end{align}
		Following \eqref{torscurvdef}, the torsion is given by
\begin{align} \label{torsion}
			\mathcal{T} = \left(\begin{array}{c||c}
				\mathcal{T}_1 & \mathcal{T}_3 \\
				\hline \hline
				\mathcal{T}_2 & 0 
			\end{array}\right)\,,
\end{align}
		where 
\bsubeq
\begin{align}
			\mathcal{T}_1 &= \left(\begin{array}{c|c} \label{t1}
				0 & \mathcal{T}_R \\
				\hline
				\mathcal{T}_L & 0  
			\end{array}\right)\,, \\
			\mathcal{T}_L &= - \text{d}F^{\dag}\text{d}G + \text{d}G^{\dag}\text{d}F - \ri\text{d}\Q^{\dag}\text{d}\Q 
			\\
			\notag 
			& \quad 
			- (F^{\dag}\text{d}G - G^{\dag}\text{d}F + \ri\Q^{\dag}\text{d}\Q)\ve^{-1}(F^{\T}\ve\text{d}F - G^{\T}\ve^{-1}\text{d}G + \Q^{\T}\text{d}\Q)
			\\
			\notag 
			& \quad 
			- (F^{\dag}\ve\text{d}\bar{F} - G^{\dag}\ve^{-1}\text{d}\bar{G}-\
			\Q^{\dag}\text{d}\bar{\Q})\ve^{-1}(F^{\dag}\text{d}G-G^{\dag}\text{d}F+\ri\Q^{\dag}\text{d}\Q)\,, 
			\\
			\mathcal{T}_R &= \ve^{-1} \bar{\mathcal{T}}_L\ve^{-1} \,, \label{tr}
			\\
			\mathcal{T}_2 &= -\text{d}D\text{d}C + \text{d}\big(DC(A^{-1})\big)\text{d}A - D\text{d}C\hat{\O} + DC(A^{-1})\text{d}A\hat{\O}
			\\
			\notag 
			& \quad
			- D^{-1}\text{d}D\big(D\text{d}C - DC(A^{-1})\text{d}A\big) 
			\\
			\notag 
			& \quad
			+ DC(A^{-1})\me^{-1}\text{d}\big((A^{-1})^{\T}C^{\T}\big)D\big(D\text{d}C - DC(A^{-1})\text{d}A\big)\,,
			\\
			\mathcal{T}_3 &= \me^{-1} \mathcal{T}_2^{\T}\,. 
\end{align}
\esubeq

The curvature is given by
\begin{align}
			\mathcal{R} = \left( \begin{array}{c|c||c}
				\mathcal{R}_1 & 0 & 0 \\
				\hline 
				0 & \mathcal{R}_2 & 0 \\
				\hline \hline 
				0 & 0 & \mathcal{R}_3
			\end{array} \right) \,,
\end{align}
		where 
\bsubeq
\begin{align}
			\mathcal{R}_1 &= \ve^{-1} \text{d}G^{\T}\ve^{-1} \text{d}G - \ve^{-1} \text{d}F^{\T}\ve\text{d}F - \ve^{-1} \text{d}\Q^{\T}\text{d}\Q 
			\\
			\notag 
			& \quad 
			-\ve^{-1} F^{\T}\ve\text{d}F\ve^{-1} F^{\T}\ve\text{d}F + \ve^{-1} F^{\T}\ve\text{d}F\ve^{-1} G^{\T}\ve^{-1} \text{d}G
			\\
			\notag 
			& \quad 
			-\ve^{-1} F^{\T}\ve\text{d}F\ve^{-1} \Q^{\T}\text{d}\Q + \ve^{-1} G^{\T}\ve^{-1} \text{d}G\ve^{-1} F^{\T}\ve\text{d}F
			\\
			\notag 
			& \quad 
			- \ve^{-1} G^{\T}\ve^{-1} \text{d}G\ve^{-1} G^{\T}\ve^{-1} \text{d}G + \ve^{-1} G^{\T}\ve^{-1} \text{d}G\ve^{-1} \Q^{\T}\text{d}\Q
			\\
			\notag 
			& \quad 
			-\ve^{-1} \Q^{\T}\text{d}\Q \ve^{-1} F^{\T} \ve\text{d}F + \ve^{-1} \Q^{\T}\text{d}\Q\ve^{-1} G^{\T}\ve^{-1} \text{d}G 
			\\
			\notag 
			& \quad 
			-\ve^{-1} \Q^{\T}\text{d}\Q\ve^{-1} \Q^{\T}\text{d}\Q\,,
			\\
			\mathcal{R}_2 &= \ve\bar{\mathcal{R}}_1\ve^{-1} \,,
			\\
			\mathcal{R}_3 &= 
			 D\text{d}C(A^{-1})\me^{-1}\text{d}(A^{-1})^{\T}C^{\T}D 
			+ \ DC\text{d}(A^{-1})\me^{-1}\text{d}(A^{-1})^{\T}C^{\T}D 
			\\
			\notag 
			& \quad 
			+ D\text{d}C(A^{-1})\me^{-1}(A^{-1})^{\T}\text{d}C^{\T}D
			+ DC\text{d}(A^{-1})\me^{-1}(A^{-1})^{\T}\text{d}C^{\T}D 
			\\
			\notag 
			& \quad 
			+ D^{-1}\text{d}C\me^{-1}C^{\T}C(A^{-1})\me^{-1}\text{d}((A^{-1})^{\T}C^{\T})D
			\\
			\notag
			&\quad 
			+ D^{-1}C\me^{-1}\text{d}C^{\T}C(A^{-1})\me^{-1}\text{d}((A^{-1})^{\T}C^{\T})D
			\\
			\notag 
			& \quad
			+ \big(DC(A^{-1})\me^{-1}\text{d}((A^{-1})^{\T}C^{\T})D\big)^2\,.
\end{align}
\esubeq
		%
%
		%
		%
\subsection{Covariant derivatives} \label{covariant derivatives}
The vierbein and connection (as well as curvature and torsion) can be decomposed into the bases corresponding to the superalgebra $\mathcal{W}$ and the algebra $\mathcal{H}$. 
Accordingly, we must introduce a basis $W_A = (P_a, q_{I\a}, \bar{q}_{I}{}^{\ad})$ for the superalgebra $\mathcal{W}$ and a basis $H_{\hat{I}} = (M_{ab}, \mathcal{J}_{IJ})$ for the algebra $\mathcal{H}$. 
The elements $\underline{h}$ of $\mathcal{H}$ and $\underline{w}$ of $\mathcal{W}$, given by \eqref{stabform} and \eqref{compel}, may be written as a linear combination of generators
\bsubeq
\begin{align} \label{stability algebra generators}
	\underline{h} &= \frac{1}{2}n^{ab}M_{ab} + \frac{1}{2}r^{IJ} \cJ_{IJ}\,,
	\\
	\underline{w} &= v^{a}P_{a} + \ri( \psi^{I\a}q_{I\a} + \bar{\psi}_{I\ad}\bar{q}_{I}{}^{\ad})\,.
\end{align}
\esubeq
Making use of \eqref{stabform} and \eqref{compel}
allows us to read off the graded commutation relations for
the generators of  $\mathcal{W}$ and $\mathcal{H}$ 
\bsubeq \label{superalgebra}
\begin{align} 
	[M_{ab},M_{cd}] &= \eta_{ad}M_{bc} - \eta_{ac}M_{bd} + \eta_{bc}M_{ad} - \eta_{bd}M_{ac}
	\,,
	\\
	[P_{a},P_{b}] &= 4 M_{ab}
	\,,
	\\
	[M_{ab},P_{c}] &= \eta_{cb}P_{a}-\eta_{ca}P_{b}
	\,,
	\\
	[M_{ab},q_{I\a}] &= -(\s_{ab})_{\a}{}^{\b}q_{I\b}
	\,,
	\\
	[M_{ab},\bar{q}_{I}{}^{\ad}] &= -(\Tilde{\s}_{ab})^{\ad}{}_{\bd}\bar{q}_{I}{}^{\bd}
	\,,
	\\
	[P_{a},q_{I\a} ] &= -\ri(\s_a)_{\a\ad}\bar{q}_{I}{}^{\ad}
	\,,
	\\
	[P_{a},\bar{q}_{I}{}^{\ad}] &= -\ri(\Tilde{\s}_{a})^{\ad\a}q_{I\a}
	\,,
	\\
	[\mathcal{J}_{IJ},\mathcal{J}_{KL}] &= \d_{JK}\mathcal{J}_{IL} - \d_{IK}\mathcal{J}_{JL} - \d_{JL}\mathcal{J}_{IK} + \d_{IL}\mathcal{J}_{JK} 
	\,,
	\\
	[M_{ab},\mathcal{J}_{IJ}] &= 0
	\,,
	\\
	[P_a,\mathcal{J}_{IJ} ] &= 0
	\,,
	\\
	\{q_{I\a},q_{J\b}\} &= -8\d_{IJ}M_{\a\b} -4  \ve_{\a\b}\mathcal{J}_{IJ}
	\,,
	\\
	\{\bar{q}_{I}{}^{\ad},\bar{q}_{J}{}^{\bd}\} &= 8\d_{IJ}\bar{M}^{\ad\bd} +4  \ve^{\ad\bd}\mathcal{J}_{IJ}
	\,,
	\\
	\{q_{I\a},\bar{q}_{J}{}^{\ad}\} &= 2\ri\d_{IJ}(\s^{a})_{\a}{}^{\ad}P_a
	\,,
	\\
	[\mathcal{J}_{IJ}, q_{K\a} ] &= \d_{JK}q_{I\a} - \d_{IK}q_{J\a}
	\,.
\end{align}
\esubeq
These relations constitute the $\mathcal{N}$-extended AdS superalgebra
$ \mathfrak{osp}(\mathcal{N}|4;\mathbb{R})_{C}$.

The vierbein and the torsion two-form, as elements of $\mathcal{W}$, can be decomposed 
with respect to the basis as
\bsubeq
\begin{align}
	\textbf{E} &= \textbf{E}^a P_a + \ri (\textbf{E}^{I\a} q_{I\a} + \bar{\textbf{E}}_{I\ad}\bar{q}^{I\ad})\,,
	\\
	\mathcal{T} &= \mathcal{T}^a P_a + \ri( \mathcal{T}^{I\a}q_{I\a}+\mathcal{T}_{I\ad}\bar{q}^{I\ad}) \,,
\end{align}
\esubeq
to obtain the one-form $\textbf{E}^{A} = (\textbf{E}^{a}, \textbf{E}^{I\a},\bar{\textbf{E}}_{I\ad})$ and the the torsion $\mathcal{T}^{A} = (\mathcal{T}^{a}, \mathcal{T}^{I\a}, \mathcal{T}_{I\ad})$.
A similar procedure follows for the curvature. We may further decompose the torsion and curvature components as
\bsubeq
\begin{align}
	\mathcal{T}^{A} &= \frac{1}{2}\textbf{E}^B\wedge\textbf{E}^C\mathcal{T}_{CB}{}^{A}\,,
	\\
	\mathcal{R} &= \frac{1}{2}\textbf{E}^B\wedge\textbf{E}^{C}\left(\frac{1}{2}\mathcal{R}_{CB}{}^{de}M_{de} + \frac{1}{2}\mathcal{R}_{CB}{}^{MN}\mathcal{J}_{MN}\right)\,. 
\end{align}
\esubeq

Building on the approach used in \cite{KT-M2007}, we can use \eqref{ewedgee} and the graded commutation relations \eqref{superalgebra} to determine the non-vanishing components of the torsion and curvature to be
\bsubeq
\begin{align} \label{nonvanishingcomps}
	\mathcal{T}_{I\a}{}^{J\bd a} &= 2\ri\d_{I}{}^{J}(\s^{a})_{\a}{}^{\bd}
	\,,
	\\
	\mathcal{T}_{a I\a J\bd} &= \ri\d_{IJ}(\s_{a})_{\a\bd}
	\,,
	\\
	\mathcal{T}_{a}^{I\ad J\b} &= \ri\d^{IJ}(\Tilde{\s}_{a})^{\ad\b}
	\,,
	\\
	\mathcal{R}_{ab}{}^{cd} &= -4(\d_{a}{}^{c}\d_{b}{}^{d}-\d_{a}{}^{d}\d_{b}{}^{c})
	\,,
	\\
	\mathcal{R}_{I\a J\b}{}^{\g\d} &= -4\d_{IJ}(\d_{\a}{}^{\g}\d_{\b}{}^{\d}+\d_{\a}{}^{\d}\d_{\b}{}^{\g})
	\,,
	\\
	\mathcal{R}_{I\a J\b}{}^{KL} &= -4 \ve_{\a\b}(\d_{I}{}^{K}\d_{J}{}^{L}-\d_{J}{}^{K}\d_{I}{}^{L})
	\,,
	\\
	\mathcal{R}^{I\ad J\bd}{}_{\gd\dd} &=  4\d^{IJ}(\d^{\ad}{}_{\gd}\d^{\bd}{}_{\dd}+\d^{\ad}{}_{\dd}\d^{\bd}{}_{\gd})
	\,,
	\\
	\mathcal{R}^{I\ad J\bd KL} &= 4\ve^{\ad\bd}(\d^{IK}\d^{JL}-\d^{JK}\d^{IL})
	\,.
\end{align}
\esubeq
These components can be used to construct the graded commutation relations of the covariant derivatives
\begin{align} \label{covdiv}
	[\mathcal{D}_A,\mathcal{D}_B\} &= - \mathcal{T}_{AB}{}^{C}\mathcal{D}_C + \frac{1}{2}\mathcal{R}_{AB}{}^{cd}M_{cd} + \frac{1}{2}\mathcal{R}_{AB}{}^{KL}\mathcal{J}_{KL}\,.
\end{align}
The algebra of covariant derivatives is thus given by
\bsubeq
\label{NowarAlgebra}
\begin{align}
	[\mathcal{D}_a,\mathcal{D}_b] &= - 4 M_{ab}
	\,,
	\\
	[\mathcal{D}_a,\mathcal{D}_{I\a}] &= -\ri(\s_{a})_{\a\bd}\bar{\mathcal{D}}_{I}{}^{\bd}
	\,,
	\\
	[\mathcal{D}_a,\bar{\mathcal{D}}_{I}{}^{\dot{\a}}] &= -\ri(\Tilde{\s}_a)^{\dot{\a}\b}\mathcal{D}_{I\b}
	\,,
	\\
	\{\mathcal{D}_{I\a},\mathcal{D}_{J\b}\} &= -8\d_{IJ}M_{\ab}  - 4\ve_{\a\b}\mathcal{J}_{IJ}
	\,,
	\\
	\{\bar{\mathcal{D}}_{I}{}^{\dot{\a}},\bar{\mathcal{D}}_{J}{}^{\dot{\b}}\} &= 
	8\d_{IJ}\bar{M}^{\ad\bd} + 4\ve^{\dot{\a}\dot{\b}}\mathcal{J}_{IJ}
	\,,
	\\
	\{\mathcal{D}_{I\a},\bar{\mathcal{D}}_{J}{}^{\dot{\b}}\} &= 
	-2\ri\d_{IJ}(\s^{a})_{\a}{}^{\dot{\b}}\mathcal{D}_{a}
	\,.
\end{align}
\esubeq
\subsection{$\N=1$ AdS superspace} \label{n=1 geometric objects}
		
		Many of the expressions in subsection \ref{geometric objects} contain $A^{-1}$ and $D$. These are, in principle, expressible in terms of $F$, $G$, and $\Q$. These expressions are, however, $\N$-dependent. Below, we will discuss both of these in the $\N=1$ case. 
		
		Using the group requirements \eqref{aka con} we can rearrange for $(A^{-1})^{\T}$
		\begin{align}
			(A^{-1})^{\T} = \me A \me^{-1}\left({\mathbbm 1} +  C^{\T}C\me^{-1}\right)^{-1}\,,
		\end{align}
		which in the $\N=1$ case yields the following expression
		\begin{align}
			(A^{-1})^{\T}  
			= \left(
			\begin{array}{c|c}
				\Tilde{F} & \ve\Bar{\Tilde{G}}\ve \\
				\hline
				\Tilde{G} & \ve^{-1}\Bar{\Tilde{F}}\ve
			\end{array}\right)
			\,,
		\end{align}
		where
		\bsubeq
		\begin{align}
			\Tilde{F} &= \ve F\ve^{-1} + \ve F \ve^{-1} \q^{\T}\q\ve^{-1} -\ve F \ve^{-1} \Q^{\T}\bar{\Q}\ve^{-1}\Q^{\dag}\Q\ve^{-1}
			\\
			\notag 
			& \quad + \ri \bar{G}\ve^{-1}\Q^{\dag}\Q\ve^{-1}
			\,,
			\\
			\Tilde{G} &= 
			-\ve^{-1} G \ve^{-1} - \ve^{-1} G \ve^{-1} \Q^{\T}\Q\ve^{-1}
			+ \ve^{-1}G\ve^{-1}\Q^{\T}\bar{\Q}\ve^{-1}\Q^{\dag}\Q\ve^{-1}
			\\
			\notag 
			& \quad 
			-\ri \bar{F} \ve^{-1}\Q^{\dag}\Q\ve^{-1}
			\,.
		\end{align}
		\esubeq
		Furthermore, $D$ has the explicit solution 
		\begin{align}
			D &= 1 - \frac{1}{2}\Q^2 - \frac{1}{2}\bar{\Q}^2 - \frac{1}{4}\Q^2\bar{\Q}^2 \,.
		\end{align}
		We can use these expressions to compute $\cE$ from the vierbein \eqref{vierbeinoddblock}. For $\N=1$ it is
		\begin{align} \label{n=1vbodd}
			\cE = \left(\begin{array}{c|c}
				2\ri(\textbf{E}_{\Q})^{\a}  & 2(\bar{\textbf{E}}_{\bar{\Q}})_{\ad}
			\end{array}\right)\,,
		\end{align}
		where 
		\begin{align} \label{n=1spinorvierbein}
			\textbf{E}_{\Q} 
			&= 
			\frac{1}{2} D \text{d} \Q
			- \frac{1}{2} (1 - \frac{1}{2}\bar{\Q}\ve^{-1}\bar{\Q})\Q\ve^{-1}(F^{\T}\ve\text{d}F-G^{\T}\ve^{-1}\text{d}G)
			\\
			\notag 
			& \quad 
			- \frac{1}{2} \ri(1 + \frac{1}{2}\Q\ve^{-1}\Q)\bar{\Q}\ve^{-1}(F^{\dag}\text{d}G - G^{\dag}\text{d}F)
			\,.
		\end{align}
		This expression coincides with \eqref{e theta block general} when considered in the $\N=1$ case. 
\subsection{North and south charts}\label{chiralgeometry}
The results of subsections \ref{geometric objects}, \ref{covariant derivatives} and \ref{n=1 geometric objects} did not make use of the freedom \eqref{equivalence relation} to fix a coordinate system. In this section we will use these results to describe the geometry in the $\N=1$ case for the north and south charts, given by two-planes of the form \eqref{chiral chart} and \eqref{south chart two planes}.  

In the north chart, the vierbein \eqref{vierbein} reads 
\begin{align} \label{chiral vierbein supermatrix}
	\textbf{E}_{\text{north}} = \left(\begin{array}{c|c||c}
		0 & \l\bar{\l}e_{\a\bd} &  
		2\ri (\eta_{\q})_{\a}\\
		\hline 
		-\l\bar{\l}{e}^{\ad\b} & 0 & - 2(\bar{\eta}_{\bar{\q}})^{\ad} \\
		 \hline \hline
		  2\ri (\eta_{\q})^{\b} &  2(\bar{\eta}_{\bar{\q}})_{\bd} & 0
	\end{array}\right)\,,
\end{align}

where $\eta_{\q}$ is computed using \eqref{n=1spinorvierbein} as
\begin{align} \label{chiralspinorvierbein}
	\eta_{\q} &= \left(\l - 2\l^{3}\q^{2} + 2\l\bar{\l}^{2}\bar{\q}^{2} + 4\l^{3}\bar{\l}^{2}\q^{2}\bar{\q}^{2} - 4\ri\l^{3}(\q x_{+} \bar{\q})\right)\text{d}\q 
	\\
	\notag
	& \quad + \left(\ri\l\bar{\l}^{2}(1+2\l^{2}\q^{2})\bar{\q}\ve^{-1} - \l^{3}(1+2\bar{\l}^{2}\bar{\q}^{2})(\q x_{+})\right)\Tilde{e}\,.  
\end{align}
In the above, $\text{d}\q$ and $\tilde{e} = e^{a}(\tilde{\s}_{a})$ 
are the flat $\N=1$ superspace vielbeins. 
The general forms for the vielbeins of a superspace with superconformally flat geometry are 
\bsubeq \label{scfvierbeins}
\begin{align}
	\textbf{E}^{a} &= \re^{-\frac{1}{2}(\s+\bar{\s})}e^{a}\,,
	\\
	(\textbf{E}_{\q})^{\a} &= \re^{\frac{1}{2}\s-\bar{\s}}\left(\text{d}\q^{\a} + \frac{\ri}{4}(\bar{D}_{\ad}\bar{\s})(\Tilde{\s}_{b})^{\ad\a}e^{b}\right) \,,
\end{align}
\esubeq
where $\s \,\, (\bar{\s})$ is chiral (antichiral). 
In our case it is straightforward to compute the coefficients in \eqref{scfvierbeins}, which yields the following expression
\begin{align}
	\label{5.37}
	\l &= \re^{-\frac{1}{2}\s}\,.
\end{align}
Indeed, \eqref{chiral vierbein supermatrix} can be shown to take the form 
\begin{align} \label{chiralvbsm2}
	\textbf{E}_{\text{north}} = \left(\begin{array}{c|c||c}
		0 & (\textbf{E}_{\text{north}})_{\a\bd} &  2\ri (\textbf{E}_{\q})_{\a} \\
		\hline 
		-({\textbf{E}}_{\text{north}})^{\ad\b} & 0 & - 2(\bar{\textbf{E}}_{\bar{\q}})^{\ad} \\
		\hline \hline
		2\ri (\textbf{E}_{\q})^{\b} & 2(\bar{\textbf{E}}_{\bar{\q}})_{\bd} & 0
	\end{array}\right)\,,
\end{align}
with $\tilde{\textbf{E}} = \textbf{E}^{a}(\tilde{\s}_{a})$ and $(\textbf{E}_{\q})$ given by \eqref{scfvierbeins}. The connection is given by
\begin{align}
	\bf{\O}_{\text{north}} = \left(\begin{array}{c|c||c}
		\O_{\a}{}^{\b} & 0 & 0 \\
		\hline 
		0 & -\bar{\O}^{\ad}{}_{\bd} & 0 \\
		\hline \hline
		0 & 0 & 0
	\end{array}\right)\,,
\end{align}
where the components of the connection read
\begin{align} \label{connection components}
	\O_{\a}{}^{\b} &= 
	e^{m}\big(\l^{-1}\d_{\a}{}^{\b}\partial_{m}\l + \l^{-1}(\tilde{\s}_{m})^{\bd\b}\partial_{\a\bd}\l\big)
	+ 
	\text{d}\q^{\g}\big(\l^{-1}\d_{\a}{}^{\b}D_{\g}\l - 2\l^{-1}\d_{\g}{}^{\b}D_{\a}\l\big)\,. 
\end{align}

We
introduce the inverse $E_{A}{}^{M}$ of the vielbein supermatrix $E_{M}{}^{A}$, 
\bsubeq \label{inverse vierbein}
\begin{align} 
\ve^{M} &= (e^{m}\,,\text{d}\q^{\a}\,, \text{d}\bar{\q}_{\ad}) = \textbf{E}^{A}E_{A}{}^{M}\,, 
\\
\textbf{E}^{A} &= (\textbf{E}^{a}\,,(\textbf{E}_{\q})^{\a}\,, (\bar{\textbf{E}}_{\bar{\q}})_{\ad}) = \ve^{M}E_{M}{}^{A}\,.
\end{align}
\esubeq 
Let us then define the vector fields 
\begin{align}
	E_{A} &=
	\big(
	E_{a}\,, 
	E_{\a}\,,
	\bar{E}^{\ad}
	\big) =
	 E_{A}{}^{M}D_{M}\,.
\end{align}
 Here $D_M :=(\partial_{m}\,, D_{\mu}\,, \bar{D}^{\dot \mu})$ are the $\N=1$ flat superspace covariant derivatives. 
 We find 
 \bsubeq
 \begin{align}
 	E_{a} &= \l^{-1}\bar{\l}^{-1}\partial_{a}  - \frac{\ri}{4}\l^{-1}\bar{\l}^{-1}(\bar{D}_{\ad}\bar{\s})(\tilde{\s}_{a})^{\ad\a}D_{\a}
 	\\
 	\notag 
 	& \quad
 	- \frac{\ri}{4}\l^{-1}\bar{\l}^{-1}(D^{a}\s)(\s_{a})_{\a\ad}\bar{D}^{\ad}\,,
 	\\
 	E_{\a} &= \l\bar{\l}^{-2}D_{\a} \,,
 	\\
 	\bar{E}^{\ad} &= \bar{\l}\l^{-2}\bar{D}^{\ad}\,. 
 \end{align}
 \esubeq

 The components of the connection $\bf{\O}_{\text{north}}$ were given with respect to the basis $\ve^{M}$ in \eqref{connection components}. 
 Using the inverse vierbein defined by \eqref{inverse vierbein}, the connection can be decomposed into the basis $\{\textbf{E}^{A}\}$,  with which we can then construct explicit expressions for the covariant derivatives
\begin{align}
	\mathcal{D}_{A} = E_{A} + \frac{1}{2}\O_{A}{}^{bc}M_{bc}
	= \big(\mathcal{D}_{a}\,, \mathcal{D}_{\a}\,, \bar{\mathcal{D}}^{\ad}\big)\,.
\end{align}
They take the following form
\bsubeq \label{north chart covariant derivatives}
\begin{align}
	\mathcal{D}_{a} &=- \frac{\rm i}{4} (\tilde{\s}_a)^{\bd \b} \big\{ \cD_\b, \bar \cD_\bd \big\} \,,
	\\
	\mathcal{D}_{\a} &= \l\bar{\l}^{-2}D_{\a} -2\bar{\l}^{-2}\big(D^{\b}\l\big)M_{\a\b} \,,
	\label{spinor derivative} \\
	\bar{\mathcal{D}}^{\ad} &= \bar{\l}\l^{-2}\bar{D}^{\ad} + 2\l^{-2}\big(\bar{D}_{\bd}\bar{\l}\big)\bar{M}^{\ad\bd} \,.
	\label{conjugate spinor derivative}
\end{align}
\esubeq
The expressions \eqref{north chart covariant derivatives} can be seen to coincide with the general form for the covariant derivatives of a conformally flat superspace. 


In the south chart, the vierbein \eqref{vierbein} is given by
\begin{align} \label{south vierbein}
	\textbf{E}_{\text{south}} &= 
	\left(\begin{array}{c|c||c}
	0 & \g\bar{\g}e'_{\a\bd} & 2\ri({\eta}_{{\xi}})_{\a} \\
	\hline 
	-\g \bar{\g} {e}'^{\ad\b} & 0 & -2(\bar{\eta}_{\bar{\xi}})^{\ad} \\
	\hline \hline
	2\ri (\eta_{\xi})^{\b} & 2(\bar{\eta}_{\bar{\xi}})_{\bd} & 0	
	\end{array}
\right)\,,
\end{align}
where $\eta_{\xi}$ is 
\begin{align}
	\eta_{\xi} &= 
	\left(\g - 2\g^{2}\xi^{2} + 2\g\bar{\g}^{2}\bar{\xi}^{2} + 4\g^{2}\bar{\g}^{2}\xi^{2}\bar{\xi}^{2} -4\ri\g^{3}(\xi y_{+} \bar{\xi})\right)\text{d}\xi 
	\\
	\notag 
	& \quad 
	+ \left(\ri \g \bar{\g}^{2}(1+2\g^{2}\xi^{2})\bar{\xi}\ve^{-1} - \g^{3}(1+2\bar{\g}^{2}\bar{\xi}^{2})(\xi\ve^{-1}y_{+}^{\T})\right)\tilde{e}' 
	\,.
\end{align}
We showed in section \ref{n s cosets} that the coset representatives in the north and south charts were related by a little group transformation, see \eqref{s coset to n coset}. 
Under such a transformation, the vierbein and connection transform as follows
\bsubeq
\begin{align}
	\textbf{E} &\rightarrow \textbf{E}' = h\textbf{E}h^{-1} \,, 
	\\
	{\bf{\O}} &\rightarrow {\bf{\O}}' = h {\bf{\O}} h^{-1} - \text{d} h h^{-1} \,. 
\end{align}
\esubeq
We can see then that the vierbein supermatrix in the north chart is related to that in the south chart by 
\begin{align}
	\textbf{E}_{\text{north}} = h \textbf{E}_{\text{south}} h^{-1}\,,
\end{align}
which yields
\bsubeq
\begin{align}
	(\textbf{E}_{\text{north}})_{\a\ad} &= (n \textbf{E}_{\text{south}} n^{\dag})_{\a\ad} \,,
	\\
	(\eta_{\q})^{\a} &= (\eta_{\xi})^{\b}(n^{-1})_{\b}{}^{\a} \,,
\end{align}
\esubeq
with $n^{-1}$ given by \eqref{n inverse block}. 
The vector fields $E_{A}$ are also related in the intersection of the two charts. We find
\bsubeq
\begin{align}
	({E_{\text{north}}})^{\ad\a} &= ((n^{\dag})^{-1} E_{\text{south}} n^{-1})^{\ad\a}
	\,,
	\\
	(E_{\text{north}})_{\a} &= n_{\a}{}^{\b}(E_{\text{south}})_{\b}\,. 
\end{align}
\esubeq


\section{Conformally flat supergeometry}
\label{section6}

This section is devoted to a description of the most general four-dimensional conformally flat supergeometry. Our approach will be to begin with a general conformally flat superspace whose local structure group is the superconformal group.\footnote{Such a supergeometry is known as a conformal superspace. They may also be employed to describe non-conformally flat supergeometries, see e.g. \cite{ButterN=1,ButterN=2} for more details.} Then, by performing a series of gauge fixings, and passing through the conventional $\sU(\cN)$ and $\sSU(\cN)$ superspaces, we realise the ${\rm AdS}$ supergeometry within this framework.   


\subsection{Conformal superspace: conformally flat geometry} \label{section6.1}

We consider a conformally flat $\N$-extended
superspace $\cM^{4|4\N}$, parametrised by local coordinates 
$z^{M} = (x^{m},\theta^{\m}_\imath,\bar \theta_{\dot{\mu}}^\imath)$, where $m=0, 1, 2, 3$, $\mu = 1, 2$, $\dot{\mu} = \dot{1}, \dot{2}$ and
$\imath = \underline{1}, \dots, \underline{\cN}$.
The structure group is chosen to be $\sSU(2,2|\cN)$, the $\N$-extended superconformal group. 
Its corresponding Lie superalgebra, $\mathfrak{su}(2,2|\cN)$, is spanned by the translation $P_A=(P_a, Q_\a^i ,\bar Q^\ad_i)$, Lorentz $M_{ab}$,  $R$-symmetry
$\mathbb{Y}$ and $\mathbb{J}^{i}{}_j$, dilatation $\mathbb{D}$,  and the special conformal $K^A=(K^a, S^\a_i ,\bar S_\ad^i)$ generators, see appendix \ref{AppendixC} for more details.
The geometry of this superspace is encoded within the conformally covariant derivatives $\nabla_A = (\nabla_a,\nabla_\a^i,\bar{\nabla}_i^\ad)$, which take the form:
\begin{align}
	\label{6.1}
	\nabla_A &= E_A{}^M \partial_M - \hf \Omega_A{}^{bc} M_{bc} - \Phi_A{}^j{}_k \mathbb{J}^{k}{}_j - \ri \Phi_A \mathbb{Y}
	- B_A \mathbb{D} - \frak{F}_{AB} K^B \eol
	&= E_A{}^M \partial_M - \Omega_A{}^{\b\g} M_{\b\g} - \bar{\Omega}_A{}^{\bd\gd} \bar{M}_{\bd\gd}
	- \Phi_A{}^j{}_k \mathbb{J}^{k}{}_j - \ri \Phi_A \mathbb{Y} - B_A \mathbb{D} - \frak{F}_{A B} K^B ~,
\end{align} 
where $E_{A}{}^M$ denotes the inverse supervielbein and the remaining superfields are connections associated with the non-translational generators of the superconformal group.

By definition, the gauge group of conformal supergravity is generated by local transformations of the form
\begin{align}
	\label{6.2}
	\nabla_A' = \re^{\mathscr{K}} \nabla_A \re^{-\mathscr{K}} ~, \qquad
	\mathscr{K} =  \xi^B \nabla_B+ \hf K^{bc} M_{bc} + \S \mathbb{D} + \ri \rho \mathbb{Y} 
	+ \chi^{i}{}_j \mathbb{J}^{j}{}_i + \L_B K^B ~ ,
\end{align}
where the gauge parameters satisfy natural reality conditions. Given a conformally covariant tensor superfield $\cU$ (with its indices suppressed), it transforms under such transformations as follows:
\begin{align}
	\label{6.3}
	\cU' = \re^{\mathscr{K}} \cU~.
\end{align}

In general, the algebra of covariant derivatives $[\nabla_A, \nabla_B\}$ should be constrained such that it: (i) has a super Yang-Mills structure; and (ii) is expressed solely in terms of a single superfield, the super-Weyl tensor. In this section, we will restrict our attention to conformally flat backgrounds, which are characterised by vanishing super-Weyl tensor. As a result, the only non-vanishing sector of $[\nabla_A, \nabla_B\}$ is
\bea
\label{6.4}
\{ \nabla_\a^i , \bar{\nabla}^{\bd}_j \} = - 2 \ri \d^i_j \nabla_\a{}^{\bd} ~.
\eea


\subsection{Degauging (i): $\sU(\cN)$ superspace}
\label{Section6.2}

According to eq. \eqref{6.2}, under an infinitesimal special superconformal gauge transformation $\mathscr{K} = \Lambda_{B} K^{B}$, the dilatation connection transforms as follows
\bea
\d_{\mathscr{K}} B_{A} = - 2 \Lambda_{A} ~.
\eea
Thus, it is possible to impose the gauge
$B_{A} = 0$, which completely fixes 
the special superconformal gauge freedom.\footnote{There is a class of residual gauge transformations which preserve this gauge. They generate the super-Weyl transformations of the degauged geometry.} As a result, the corresponding connection is no longer required for the covariance of $\nabla_A$ under the residual gauge freedom and
may be
extracted from $\nabla_{A}$,
\bea
\nabla_{A} &=& \cD_{A} - \mathfrak{F}_{AB} K^{B} ~. \label{ND}
\eea
Here the operator $\cD_{A} $ involves only the Lorentz and $R$-symmetry connections
\bea
\cD_A = E_A{}^M \partial_M - \frac{1}{2} \O_A{}^{bc} M_{bc} - \Phi_A{}^j{}_k \mathbb{J}^{k}{}_j - \ri \F_A \mathbb{Y}~.
\eea

The next step is to relate the special superconformal connection
$\mathfrak{F}_{AB}$  to the torsion tensor associated with $\cD_A$. To do this, one can  make use of the relation
\bea
\label{4.3}
[ \cD_{A} , \cD_{B} \} &=&  [ \nabla_{A} , \nabla_{B} \} + \big(\cD_{A} \mathfrak{F}_{BC} - (-1)^{\e_A \e_B} \cD_{B} \mathfrak{F}_{AC} \big) K^C + \mathfrak{F}_{AC} [ K^{C} , \nabla_B \} \non \\
&& - (-1)^{\e_A \e_B} \mathfrak{F}_{BC} [ K^{C} , \nabla_A \} - (-1)^{\e_B \e_C} \mathfrak{F}_{AC} \mathfrak{F}_{BD} [K^D , K^C \} ~.
\eea
In conjunction with \eqref{6.4}, this relation leads to a set of consistency conditions that are equivalent to the Bianchi identities of (conformally flat) $\sU(\cN)$ superspace \cite{Howe}. 
Their solution expresses the components of $\mathfrak{F}_{AB}$ in terms of the torsion 
tensor of $\sU(\cN)$ superspace and completely determines the algebra $[ \cD_{A} , \cD_{B} \}$. 


\subsubsection{$\cN=1$ case}

We begin by solving eq. \eqref{4.3} in the $\cN=1$ case. The outcome of this analysis is:
\begin{subequations} \label{N=1connections}
	\bea
	\mathfrak{F}_{\a \b} & = & \frac{1}{2} \ve_{\a \b} \bar{R} ~, \quad {\mathfrak{F}}_{\ad \bd} = -\frac{1}{2} \ve_{\ad \bd} R ~, \quad
	\mathfrak{F}_{\a \bd} 
	= -\bar{\mathfrak{F}}_{\bd \a} 
	= \frac{1}{4} G_{\a \bd} ~, 
	\\
	\mathfrak{F}_{\a , \b \bd} & = & - \frac{\ri}{4} \cD_{\a} G_{\b \bd} - \frac{\ri}{6} \ve_{\a \b} \bar{X}_{\bd}  = \mathfrak{F}_{\b \bd , \a} ~, \\
	{\mathfrak{F}}_{\ad , \b \bd} &=& \frac{\ri}{4} \bar{\cD}_{\ad} G_{\b \bd} + \frac{\ri}{6} \ve_{\ad \bd} X_{\b} =\mathfrak{F}_{\b \bd , \ad} ~, \\
	\mathfrak{F}_{\a \ad , \b \bd} & = & - \frac{1}{8} \big[ \cD_{\a} , \bar{\cD}_{\ad} \big] G_{\b \bd} - \frac{1}{12} \ve_{\ad \bd} \cD_{\a} X_{\b} + \frac{1}{12} \ve_{\a \b} \bar{\cD}_{\ad} \bar{X}_{\bd}  \non \\
	&& 
	+ \frac{1}{2} \ve_{\a \b} \ve_{\ad \bd} \bar{R} R
	+ \frac{1}{8} G_{\a \bd} G_{\b \ad} ~.
	\eea
\end{subequations}
Here $R$ is a chiral scalar superfield
\begin{subequations}
	\label{N=1BIs}
	\bea
	\bar{\cD}_{\ad} R = 0 ~, \qquad \mathbb{Y} R = -2 R~, 
	\eea
while $X_\a$ is the chiral field strength of a $\sU(1)$ vector multiplet
	\bea
	\bar{\cD}_{\ad} X_{\a} = 0 ~, \qquad \cD^\a X_\a = \bar{\cD}_\ad \bar{X}^\ad~, \qquad \mathbb{Y} X_\a = - X_\a~,
	\eea
and $G_{\a \ad}$ is a real vector superfield. These are related via
	\bea
	X_{\a} &=& \cD_{\a}R - \bar{\cD}^{\ad}G_{\a \ad} ~, \label{Bianchi1} \\
	{\rm i} \cD_{(\a}{}^{\gd} G_{\b ) \gd} &=&  \frac{1}{3} \cD_{(\a} X_{\b)} ~.
	\eea
\end{subequations}

We now pause and comment on the geometry described by $\cD_A$. In particular, by employing \eqref{4.3} one arrives at the following anti-commutation relation
\be
\label{4.6}
\{ \cD_\a , \bar{\cD}_{\ad} \} = - 2 \ri \cD_{\aa} - G^{\b}{}_{\ad} M_{\a \b}
+ G_\a{}^{\bd} \bar{M}_{\ad \bd} + \frac{3}{2} G_\aa \mathbb{Y} ~.
\ee
It follows that if one performs the shift
\begin{subequations}
	\label{A.16}
	\bea
	\cD_{\a \ad} \longrightarrow \cD_{\a \ad} + \frac{\ri}{2} G^{\b}{}_{\ad} M_{\a \b} &-& \frac{\ri}{2} G_{\a}{
	}^{\bd} \bar{M}_{\ad \bd} - \frac{3 \ri}{4} G_{\a \ad} \mathbb{Y} ~, 
	\eea
\end{subequations} 
then the $G$-dependent terms in \eqref{4.6} vanish.
The resulting algebra of covariant derivatives, up to dimension-$3/2$, takes the form
\begin{subequations} \label{U(1)algebra}
	\bea
	\{ \cD_{\a}, \cD_{\b} \} &=& -4{\bar R} M_{\a \b}~, \qquad
	\{\cDB_{\ad}, \cDB_{\bd} \} =  4R {\bar M}_{\ad \bd}~, \label{U(1)algebra.a}\\
	&& {} \qquad \{ \cD_{\a} , \cDB_{\ad} \} = -2{\rm i} \cD_{\a \ad} ~, 
	\label{U(1)algebra.b}	\\
	\big[ \cD_{\a} , \cD_{ \b \bd } \big]
	& = &
	{\rm i}
	{\ve}_{\a \b}
	\Big({\bar R}\,\cDB_\bd + G^\g{}_\bd \cD_\g
	- (\cD^\g G^\d{}_\bd)  M_{\g \d}
	\Big)
	+ {\rm i} (\cDB_{\bd} {\bar R})  M_{\a \b}
	\non \\
	&&
	-\frac{\ri}{3} \ve_{\a\b} \bar X^\gd \bar M_{\gd \bd} - \frac{\ri}{2} \ve_{\a\b} \bar X_\bd \mathbb{Y}
	~, \label{U(1)algebra.c}\\
	\big[ {\bar \cD}_{\ad} , \cD_{\b\bd} \big]
	& = &
	- {\rm i}
	\ve_{\ad\bd}
	\Big({R}\,\cD_{\b} + G_\b{}^\gd \cDB_\gd
	- (\cDB^{\gd} G_{\b}{}^{\dd})  \bar M_{\gd \dd}
	\Big) 
	- {\rm i} (\cD_\b R)  {\bar M}_{\ad \bd}
	\non \\
	&&
	+\frac{\ri}{3} \ve_{\ad \bd} X^{\g} M_{\g \b} - \frac{\ri}{2} \ve_{\ad\bd} X_\b \mathbb{Y}
	~, \label{U(1)algebra.d}
	\eea
\end{subequations}
which describes a $\sU(1)$ superspace \cite{Howe,GGRS} with vanishing super-Weyl tensor.

Above we made use of the special conformal gauge freedom to degauge from conformal to $\sU(1)$ superspace. Now, we will show that the residual dilatation symmetry manifests in the latter as super-Weyl transformations. To preserve the gauge $B_{A}=0$, every local dilatation transformation should be accompanied by a compensating special conformal one
\begin{align}
	\mathscr{K}(\S) = \S \mathbb{D} + \L_{B} (\S)K^{B} \quad \implies \quad B'_A = 0~.
\end{align}
This is the case only if the special conformal parameter is
\bea
\L_{A}(\S) = \hf \nabla_A \S~.
\eea
	
We now determine what transformation of $\cD_A$ and the torsions of $\sU(1)$ superspace this induces. They may be determined by making use of the following relation
\bea
\nabla'_{A} &=& \cD'_{A} - \mathfrak{F}'_{AB} K^{B} = \re^{\mathscr{K}(\S)} \nabla_A \re^{-\mathscr{K}(\S)}~.
\eea
Specifically, one finds that the super-Weyl transformations of the degauged geometry are:
\begin{subequations}\label{FinitesuperWeylTf}
	\bea
	\cD_{\a}' & = & \re^{\frac{1}{2} \S} \left( \cD_{\a} + 2 \cD^{\b} \Sigma M_{\b \a} - \frac{3}{2} \cD_{\a} \Sigma \mathbb{Y} \right) ~, \\
	\cDB_{\ad}' & = & \re^{\frac{1}{2} \S} \left( \cDB_{\ad} + 2 \cDB^{\bd} \S {\bar M}_{\bd \ad} + \frac{3}{2} \cDB_{\ad} \S \mathbb{Y} \right) ~, \\
	\cD_{\a \ad}' & = & \re^{\S} \Big( \cD_{\a \ad} + {\rm i} \cD_{\a} \S \cDB_{\ad} + {\rm i} \cDB_{\ad} \S \cD_{\a} + {\rm i} \left( \cDB_{\ad} \cD^{\b} \S  + 2 \cDB_{\ad} \S \cD^{\b} \S \right) M_{\b \a} \non \\ && + {\rm i} \left( \cD_{\a} \cDB^{\bd} \S + 2 \cD_{\a} \S \cDB^{\bd} \S \right) { \bar M}_{\bd \ad} + 3{\rm i} \Big( \frac{1}{4} \left[ \cD_{\a} , \cDB_{\ad} \right] \S +  \cD_{\a} \S \cDB_{\ad} \S \Big) \mathbb{Y} \Big) ~.~~~ \\
	R' & = & \re^{\S} \Big( R + \frac{1}{2} \cDB^{2} \S - ( \cDB \S )^{2} \Big) ~, \\
	G_{\a \ad}' & = &  \re^{\S} \Big( G_{\a \ad} + [ \cD_{\a} , \cDB_{\ad} ] \S + 2 \cD_{\a} \S \cDB_{\ad} \S \Big) ~, \\
	X_{\a}' & = & \re^{\frac{3}{2} \S} \Big( X_{\a} 
	- \frac{3}{2} (\cDB^{2} - 4 R) \cD_{\a} \S \Big) ~,
	\label{swx}
	\eea
\end{subequations}
which are in agreement with the ones presented in \cite{KR19}. Additionally, for infinitesimal $\S$, these transformations may be obtained from the ones presented in \cite{Howe}.

\subsubsection{$\cN>1$ case}
We now extend the analysis presented above to the $\cN>1$ case. A routine calculation leads to the following expressions for the degauged special conformal connection:
\begin{subequations} \label{connections}
	\bea
	\mathfrak{F}_\a^i{}_\b^j  
	&=&
	-\hf\ve_{\a\b}S^{ij}
	- Y_{\a\b}^{ij}
	~,
	\qquad
	\mathfrak{F}^\ad_i{}^\bd_j  
	=
	-\hf\ve^{\ad\bd}\bar{S}_{ij}
	-\bar{Y}^{\ad\bd}_{ij}
	~,\\
	\mathfrak{F}_\a^i{}^\bd_j
	&=&
	- \mathfrak{F}^\bd_j{}_\a^i
	=
	-\d^i_jG_\a{}^\bd
	-\ri G_\a{}^\bd{}^i{}_j
	~,
	\\
	\mathfrak{F}_{\a}^{i}{}_{,\bb}
	&=&  \ri \cD_{(\a}^i G_{\b) \bd} - \frac{1}{\cN+1} \cD_{(\a}^j G_{\b) \bd}{}^i{}_j - \hf \ve_{\a \b} \Big ( \ri \cD_\g^i G^\g{}_\bd + \frac{1}{\cN-1} \cD_\g^j G^\g{}_\bd{}^i{}_j \Big ) = \mathfrak{F}_{\bb}{}_{,\a}^i ~, ~~~
	\\
	\mathfrak{F}_{\ad i ,\bb}
	&=&
	- \ri \bar{\cD}_{(\ad i} G_{\b \bd)} + \frac{1}{\cN+1} \bar{\cD}_{(\ad j} G_{\b) \bd}{}^i{}_j - \hf \ve_{\a \b} \Big ( \ri \bar{\cD}^\gd_{ i} G_{\b \gd} + \frac{1}{\cN-1} \bar{\cD}_{j}^\gd G_{\b \gd}{}^j{}_i \Big )
	=
	\mathfrak{F}_{\bb, \ad i}
	~, ~~~~~~~~~~ \\
	\mathfrak{F}_{\aa,\bb} &=& \frac{\ri}{2\cN} \Big( 
	\cD_\a^i \mathfrak{F}_{\ad i,\bb} + \bar{\cD}_{\ad i} \mathfrak{F}^i_{\a,\bb} + 2 \ri \Big( \mathfrak{F}_\a^i{}_\b^j \mathfrak{F}_{\ad i , \bd j}
	+ \mathfrak{F}^i_{\a, \bd j} \mathfrak{F}_{\ad i, \b}{}^j
	\Big)
	\Big)~.
	\eea
\end{subequations}
The dimension-1 superfields introduced above have the following symmetry properties:  
\bea
S^{ij}=S^{ji}~, \qquad Y_{\a\b}^{ij}=Y_{\b\a}^{ij}=-Y_{\a \b}^{ji}~, \qquad {G_{\a \ad}{}^{i}{}_i} = 0~,
\eea
and satisfy the reality conditions
\bea
\overline{S^{ij}} =  \bar{S}_{ij}~,\quad
\overline{Y_{\a\b}^{ij}} = \bar{Y}_{\ad\bd ij}~,\quad
\overline{G_{\b\ad}} = G_{\a\bd}~,\quad
\overline{G_{\b\ad}{}^{i}{}_j} = - G_{\a\bd}{}^j{}_{i}
~.~~~~~~
\eea
The ${\sU}(1)_R$ charges of the complex fields are:
\bea
{\mathbb Y} S^{ij}=2S^{ij}~,\qquad
{\mathbb Y}  Y_{\a\b}=2Y^{ij}_{\a\b}~.
\eea
Now, by employing \eqref{4.3}, we find that the anti-commutation relations for the spinor covariant derivatives are:
\begin{subequations} \label{U(2)algebra}
	\bea
	\{ \cD_\a^i , \cD_\b^j \}
	&=&
	4 S^{ij}  M_{\a\b} 
	+4\ve_{\a\b} Y^{ij}_{\g\d}  M^{\g\d}  
	-4\ve_{\a \b} S^{k[i} \mathbb{J}^{j]}{}_k
	+ 8{Y}_{\a\b}^{k(i}  \mathbb{J}^{j)}{}_k
	~,
	\label{U(2)algebra.a}
	\\
	\{ \cD_\a^i , \bar{\cD}^\bd_j \}
	&=&
	- 2 \ri \d_j^i\cD_\a{}^\bd
	+4\Big(
	\d^i_jG^{\g\bd}
	+\ri G^{\g\bd}{}^i{}_j
	\Big) 
	M_{\a\g} 
	+4\Big(
	\d^i_jG_{\a\gd}
	+\ri G_{\a\gd}{}^i{}_j
	\Big)  
	\bar{M}^{\bd\gd}
	\non\\
	&&
	+8 G_\a{}^\bd \mathbb{J}^i{}_j
	+4\ri\d^i_j G_\a{}^\bd{}^{k}{}_j \mathbb{J}^i{}_{k}
	+\frac{2(\cN-4)}{\cN}\Big(
	\d^i_jG_\a{}^\bd
	+\ri G_\a{}^\bd{}^i{}_j
	\Big)
	\mathbb{Y} 
	~.
	\label{U(2)algebra.b}
	\eea
	\esubeq
At the same time, the consistency conditions arising from solving \eqref{4.3} lead to the Bianchi identities:
	\begin{subequations}\label{BI-U2}
		\bea
		\cD_{\a}^{(i}S^{jk)}&=&0~, \quad
		\bar{\cD}_{\ad i}S^{jk} - \ri\cD^{\b (j}G_{\b\ad}{}^{k)}{}_i = \frac{1}{\cN+1} \d_i^{(j} \Big( 2 \bar{\cD}_{\ad l} S^{k)l} - \ri {\cD}^{\b |l|} G_{\b \ad}{}^{k)}{}_l\Big) ~, ~~~
		\\
		\cD_{(\a}^{(i}Y_{\b\g)}^{j)k}&=&0~, \quad
		\cD^{\b k} Y_{\a \b}^{ij} = - \cD_\a^{[i} S^{j]k}~, \quad
		\bar{\cD}_j^\bd Y_{\a \b}^{ij} = 2 \cD_{(\a}^i G_{\b)}{}^{\bd} - \ri \frac{\cN-2}{\cN+1} \cD_{(\a}^j G_{\b)}{}^{\bd i}{}_j~,~~~ \\
		\cD_{(\a}^{(i}G_{\b)\bd}{}^{j)}{}_k&=&\frac{1}{\cN+1} \cD_{(\a}^l G_{\b) \bd}{}^{(i}{}_l \d^{j)}_k~, 
		\qquad
		\cD_{(\a}^{[i}G_{\b)\bd}{}^{j]}{}_k=-\frac{1}{\cN-1} \cD_{(\a}^l G_{\b) \bd}{}^{[i}{}_l \d^{j]}_k~, 
		\\
		\cD_\a^iG^{\a \bd}&=&
		\frac{\ri}{2(\cN+1)} \Big( \frac{\cN+2}{\cN-1} \cD_\a^j G^{\a \bd i}{}_j + \ri \bar{\cD}^{\bd}_j S^{ij} \Big) ~.
		\eea
	\end{subequations}
	

Now, in complete analogy with the $\cN=1$ story described above, we show how the residual dilatation symmetry of conformal superspace manifests in the present geometry as super-Weyl transformations. It may be shown that the following combined dilatation and special conformal transformation, parametrised by a dimensionless real scalar superfield $\S$ = $\bar{\S}$, preserves the gauge $B_A = 0$:
\begin{align}
	\mathscr{K}(\S) = \S \mathbb{D} + \hf \nabla_B \S K^{B} \quad \implies \quad B'_A = 0~.
\end{align}
At the level of the degauged geometry, this induces the following super-Weyl transformations
\begin{subequations}
	\label{6.17}
\bea
	\cD^{' i}_\a&=&\re^{\hf \S} \Big( \cD_\a^i+2\cD^{\b i}\S M_{\a \b} + 2 \cD_{\a}^j \S \mathbb{J}^{j}{}_i 
	+ \frac{\cN-4}{2\cN} \cD_\a^i\S {\mathbb Y} \Big)
	~,
	\label{Finite_D}
	\\
	\bar{\cD}_{i}^{' \ad}&=&\re^{\hf \S} \Big( \bar{\cD}^\ad_i+2 \bar{\cD}_{i}^{\bd} \S \bar{M}_{\ad \bd} - 2 \bar{\cD}^{\ad}_j \S \mathbb{J}^{j}{}_i 
	- \frac{\cN-4}{2\cN} \bar{\cD}_{i}^{\ad} {\mathbb Y} \Big)
	~,
	\label{Finite_DB}\\
	{\cD}'_\aa&=&\re^{\S} \Big( \cD_{\a \ad} + {\rm i} \cD^i_{\a} \S \cDB_{\ad i} + {\rm i} \cDB_{\ad i} \S \cD_{\a}^i + \Big( \cD^\b{}_\ad \S - \ri \cD^{\b i} \S \cDB_{\ad i} \S \Big) M_{\a \b} \non \\ && + \Big( \cD_{\a}{}^\bd \S + \ri \cD_{\a}^i \S \cDB^{\bd}_i \S \Big) { \bar M}_{\ad \bd} 
	- 2\ri \cD_\a^i \S \cDB_{\ad j} \S \mathbb{J}^j{}_i
	- \frac{\ri(\cN-4)}{2\cN} \cD_{\a}^i \S \cDB_{\ad i} \S \mathbb{Y} 
	\Big)
	~,
	\label{Finite_DBB}
	\\
	S^{' ij}&=& \re^{\S} \Big( S^{ij}
	-\hf \cD^{ij} \S + \cD^{\a (i} \S \cD_\a^{j)} \S \Big)
	\label{Finite_S}~,
	\\
	Y^{' ij}_{\a\b}&=& \re^{\S} \Big( Y_{\a\b}^{ij}
	-\hf \cD_\a^{[i} \cD_\b^{j]} \S - \cD_\a^{[i} \S \cD_\b^{j]} \S \Big)
	\label{Finite_Y}~,
	\\
	G'_{\a\ad}&=&
	\re^{\S} \Big( G_{\a\ad}
	-{\frac{1}{4\cN}}[\cD_\a^i,\bar{\cD}_{\ad i}]\S
	-\frac{1}{2} \cD_\a^i \S \bar{\cD}_{\ad i} \S \Big)
	~,
	\label{Finite_G}
	\\
	G'_{\a\ad}{}^{i}{}_j&=& \re^{\S} \Big( G_{\a\ad}{}^{i}{}_j
	+{\frac\ri 4} \Big( [\cD_\a^{i},\bar{\cD}_{\ad j}] - \frac{1}{\cN} \d^i_j [\cD_\a^{k},\bar{\cD}_{\ad k}] \Big) \S \Big)
	~,
	\label{Finite_Gij}
	\eea
\end{subequations}
where we have made the definitions:
\begin{align}
	\cD^{ij} = \cD^{\a (i} \cD_{\a}^{j)} ~, \qquad \bar{\cD}_{i j} = \bar{\cD}_{\ad (i}\bar{\cD}^{\ad}_{j)}~.
\end{align}
In the infinitesimal case, these transformations are a special case of the ones presented in \cite{Howe}.\footnote{Recently, the super-Weyl transformations of $\cN$-extended superspace have been computed within a local supertwistor formulation approach, see \cite{HL2} for more details.} Further, for $\cN=2$, these may be read off from the finite super-Weyl transformations presented in \cite{KLRT-M2}.

	
\subsection{Degauging (ii): $\sSU(\cN)$ superspace}

In the preceeding subsection we have shown that the degauging of the $\cN$-extended conformally flat supergeometry described in section \ref{section6.1} leads to (conformally flat) $\sU(\cN)$ superspace. The latter is characterised by the property that its local structure group is $\sSL(2,\mathbb{C}) \times \sU(\cN)_R$. In the present section we will further degauge this geometry by breaking the local $R$-symmetry group down to $\sSU(\cN)_R$. 

This procedure consists of the following steps. First, one must eliminate the $\sU(1)_R$ curvature. This involves redefining $\cD_A$ to absorb such terms in the algebra of covariant derivatives and employing super-Weyl transformations to set the remaining contributions, which describe purely gauge degrees of freedom, to zero. For $\cN=1$, this role is played by the chiral spinor $X_\a$, while in the $\cN>1$ case, $G_{\aa}{}^i{}_j$ should be gauged away. Next, by performing some local $\sU(1)_R$ transformation one may always set $\F_A = 0$, and so the local $R$-symmetry group has been reduced to $\sSU(\cN)_R$. Finally, one must identify the class of residual combined super-Weyl and local $\sU(1)_R$ transformations preserving this geometry. As will be shown below, such transformations are parametrised by a dimensionless chiral scalar $\S$ (and its conjugate).

\subsubsection{$\cN = 1$ case}
As pointed out above, the spinor $X_\a$ is the chiral field strength 
of an Abelian vector multiplet and describes purely gauge degrees of freedom. By employing the super-Weyl transformatons \eqref{swx}
it is possible to fix the gauge
\bea
X_\a =0~.
\eea
By inspecting the algebra of covariant derivatives \eqref{U(1)algebra}, it is clear this leads to vanishing $\sU(1)_{R}$ curvature. Hence, in this gauge the $\sU(1)_R$ connection $\F_A$ may also be gauged away
\bea
{\F}_A =0~.
\eea
Then, the algebra of covariant derivatives \eqref{U(1)algebra} reduces to
\begin{subequations} \label{GWZalgebra}
	\bea
	\{ \cD_{\a}, \cD_{\b} \} &=& -4{\bar R} M_{\a \b}~, \qquad
	\{\cDB_{\ad}, \cDB_{\bd} \} =  4R {\bar M}_{\ad \bd}~, \\
	&& {} \qquad \{ \cD_{\a} , \cDB_{\ad} \} = -2{\rm i} \cD_{\a \ad} ~, 
	\\
	\big[ \cD_{\a} , \cD_{ \b \bd } \big]
	& = &
	{\rm i}
	{\ve}_{\a \b}
	\Big({\bar R}\,\cDB_\bd + G^\g{}_\bd \cD_\g
	- (\cD^\g G^\d{}_\bd)  M_{\g \d}
	\Big)
	+ {\rm i} (\cDB_{\bd} {\bar R})  M_{\a \b}~,\\
	\big[ {\bar \cD}_{\ad} , \cD_{\b\bd} \big]
	& = &
	- {\rm i}
	\ve_{\ad\bd}
	\Big({R}\,\cD_{\b} + G_\b{}^\gd \cDB_\gd
	- (\cDB^{\gd} G_{\b}{}^{\dd})  \bar M_{\gd \dd}
	\Big) 
	- {\rm i} (\cD_\b R)  {\bar M}_{\ad \bd}~,
	\eea
\end{subequations}
which describes a conformally flat GWZ geometry \cite{GWZ}. This algebra should be accompanied by the constraints \eqref{N=1BIs}, provided one sets $X_\a = 0$.

Equation \eqref{swx}  tells us that imposing the condition $X_\a=0$ does not fix completely the super-Weyl freedom. The residual transformations are generated 
by parameters of the form 
\bea 
\S=\hf \big(\s +\bar \s \big) ~, \qquad \bar \cD_\ad \s =0~, \qquad \mathbb{Y} \s = 0~.
\label{2.18}
\eea
However, in order to preserve the  $\sU(1)_{R}$ gauge ${\F}_A=0$, 
every residual super-Weyl transformation \eqref{2.18} must be accompanied by the following compensating $\sU(1)_{R}$ transformation
\be
\cD_A \longrightarrow \re^{-\frac{3}{4} (\s - \bar{\s}) \mathbb{Y}} \cD_A \re^{\frac{3}{4} (\s - \bar{\s}) \mathbb{Y}}~.
\ee
This leads to the transformations:
\begin{subequations} 
	\label{superweylGWZ}
	\bea
	\cD'_\a &=& \re^{ {\bar \s} - \hf \s} \Big(  \cD_\a + \cD^\b \s \, M_{\a \b} \Big) ~, \\
	\bar \cD'_\ad & = & \re^{  \s -  \hf {\bar \s}} \Big(\bar \cD_\ad +  \bar \cD^\bd  {\bar \s}  {\bar M}_{\ad \bd} \Big) ~,\\
	\cD'_{\a\ad} &=& \re^{\hf \s + \hf \bar \s} \Big( \cD_{\a\ad} 
	+\frac{\ri}{2} \bar \cD_\ad \bar \s \cD_\a + \frac{\ri}{2} \cD_\a  \s \bar \cD_\ad  
	+ \Big( \cD^\b{}_\ad \s + \frac{\ri}{2} \bar{\cD}_\ad \bar{\s} \cD^\b \s \Big) M_{\a\b} \non \\
	&& \qquad + \Big( \cD_\a{}^\bd \bar \s + \frac{\ri}{2} \cD_\a \s \bar{\cD}^\bd \bar{\s} \Big) \bar M_{\ad \bd} \Big)~, \\
	R' &=& \re^{2\s - \bar{\s}} \Big( R +\frac{1}{4} \bar{\cD}^2 \bar{\s} - \frac{1}{4} (\bar{\cD} \bar{\s})^2 \Big) ~, \\
	G_{\a\ad}'= &=& \re^{\hf \s + \hf \bar \s} \Big( G_{\a\ad} +\ri \cD_{\a\ad} ( \s - \bar \s) + \hf \cD_\a \s \bar{\cD}_\ad \bar{\s} \Big) ~.
	\eea
\end{subequations} 
In the infinitesimal limit, these transformations may be obtained from the ones given in \cite{HT}.

\subsubsection{$\cN > 1$ case}
As discussed above, in the $\cN > 1$ case, the torsion $G_\aa{}^i{}_j$ describes purely gauge degrees of freedom.
Thus, by employing the super-Weyl freedom described by eq. \eqref{6.17}, it may be gauged away
\bea
G_{\aa}{}^{i}{}_j=0~.
\label{G2}
\eea
In this gauge, it is natural to shift $\cD_a$ as follows:
\bea
\label{6.19}
\cD_a \longrightarrow \cD_a+ \frac{\ri(\cN-4)}{\cN} G_a {\mathbb Y}~.
\eea
Then, by making use of \eqref{U(2)algebra}, we find that these covariant derivatives obey the algebra:
\begin{subequations} 
	\label{4.21}
	\bea
	\{\cD_\a^i,\cD_\b^j\}&=&
	4 S^{ij}  M_{\a\b} 
	+4\ve_{\a\b} Y^{ij}_{\g\d}  M^{\g\d}  
	-4\ve_{\a \b} S^{k[i} \mathbb{J}^{j]}{}_k
	+ 8{Y}_{\a\b}^{k(i}  \mathbb{J}^{j)}{}_k~,
	\label{acr1} \\
	\{\cD_\a^i,\cDB^\bd_j\}&=&
	-2\ri\d^i_j \cD_\a{}^\bd
	+4\d^{i}_{j}G^{\d\bd}M_{\a\d}
	+4\d^{i}_{j}G_{\a\gd}\bar{M}^{\gd\bd}
	+8 G_\a{}^\bd \mathbb{J}^{i}{}_{j}~.~~~~~~~~~
	\eea
\end{subequations}
In the $\cN=2$ case this algebra of covariant derivatives coincides with conformally flat limit of the one derived by Grimm \cite{Grimm}. It should be pointed out, however, that no discussion of super-Weyl transformations was given in \cite{Grimm}. As a result, the setup of \cite{Grimm} is insufficient to describe conformal supergravity. These transformations were later computed in \cite{KLRT-M1}.

The geometric superfields appearing above obey the Bianchi identities \eqref{BI-U2}  (upon imposing \eqref{G2}). Now, by examining equations \eqref{4.21}, we see that the $\sU(1)_R$ curvature has been eliminated and therefore the corresponding connection 
is flat. Consequently, it may be set to zero via an appropriate local $\sU(1)_R$ transformation; $\F_A =0$. As a result, the gauge group reduces to $\sSL( 2, {\mathbb C}) \times \sSU(\cN)_R$. Hence, we will refer to this supergeometry as conformally flat $\sSU(\cN)$ superspace.

It turns out that the gauge conditions \eqref{G2} and $\F_A=0$ allow for residual super-Weyl transformations, which are described
by a parameter $\s$ constrained by
\be
\Big( [\cD_\a^{i},\bar{\cD}_{\ad j}] - \frac 1 \cN \d^i_j [\cD_\a^{k},\bar{\cD}_{\ad k}] \Big) \s=0~.
\label{Ucon}
\ee
The general solution of this condition is
\bea
\S = \frac{1}{2} (\s+\bar{\s})~, \qquad \bar{\cD}^\ad_i \s =0~,
\qquad {\mathbb Y} \s =0~,
\eea
where the parameter $\s$ is covariantly chiral, with zero $\sU(1)_R$ charge, but otherwise arbitrary.
To preserve the gauge condition $\F_A=0$, every super-Weyl transformation, eq. \eqref{6.17}, must be accompanied by the following compensating $\sU(1)_R$ transformation 
\be
\cD_A \longrightarrow \re^{\frac{\cN - 4}{4\cN} (\s - \bar{\s}) \mathbb{Y}} \cD_A \re^{- \frac{\cN - 4}{4\cN} (\s - \bar{\s}) \mathbb{Y}}~.
\ee
As a result, the algebra of covariant derivatives of (conformally flat) $\sSU(\cN)$ superspace is preserved by the following set of super-Weyl transformations:
\begin{subequations}
	\label{SU(N)sW}
	\begin{align}
	\cD_\a^{'i}&= \re^{\frac{\cN-2}{2\cN} \s + \frac 1 \cN \bar{\s}} \Big( \cD_\a^i+ \cD^{\b i}\s M_{\a \b} + \cD_{\a}^j\s \mathbb{J}^{i}{}_j \Big) ~, 
	\\ 
	\bar{\cD}_{i}^{' \ad}&=\re^{\frac{1}{\cN} \s + \frac{\cN-2}{2\cN} \bar{\s}} \Big( \bar{\cD}^\ad_i-\bar{\cD}_{ \bd i} \bar{\s} \bar{M}^{\ad \bd} - \bar{\cD}^{\ad}_j \bar{\s} \mathbb{J}^{j}{}_i \Big)~,
	\\
	\cD_\aa' &= \re^{\hf \s + \hf \bar{\s}} \Big(\cD_\aa + \frac{\rm i}{2} \cD^i_{\a} \s \cDB_{\ad i} + \frac{\rm i}{2} \cDB_{\ad i} \bar{\s} \cD_{\a}^i + \hf \Big( \cD^\b{}_\ad (\s + \bar \s ) - \frac{\ri}{2} \cD^{\b i} \s \cDB_{\ad i} \bar{\s} \Big) M_{\a \b} \non \\ & + \hf \Big( \cD_{\a}{}^\bd (\s + \bar{\s}) + \frac{\ri}{2} \cD_{\a}^i \s \cDB^{\bd}_i \bar{\s} \Big) { \bar M}_{\ad \bd} 
	- \frac \ri 2 \cD_\a^i \s \cDB_{\ad j} \bar{\s} \mathbb{J}^j{}_i \Big) ~, \\
	S^{' ij}&= \re^{\frac{\cN-2}{\cN} \s + \frac 2 \cN \bar{\s}} \Big( S^{ij}-{\frac14}\cD^{ij} \s + \frac 1 4 \cD^{\a (i} \s \cD_\a^{j)} \s \Big)~, 
	\label{super-Weyl-S} \\
	Y^{' ij}_{\a\b}&= \re^{\frac{\cN-2}{\cN} \s + \frac 2 \cN \bar{\s}} \Big( Y_{\a\b}^{ij}-{\frac14}\cD_\a^{[i} \cD_\b^{j]}\s - \frac14 \cD_\a^{[i} \s \cD_\b^{j]} \s \Big)~,
	\label{super-Weyl-Y} \\
	G_{\aa}' &=
	\re^{\frac12 \s + \frac12 \bar{\s}} \Big(
	G_{\a\bd} -{\frac{\ri}4}
	\cD_{\a \ad} (\s-\bar{\s})
	-\frac{1}{8} \cD_\a^i \s \bar{\cD}_{\ad i} \bar{\s}
	\Big)
	~.
	\label{super-Weyl-G}
	\end{align}
\end{subequations}
For $\cN=2$ case these transformations are a special case of the ones given in \cite{KLRT-M1}. It is important to point out that for $\cN=4$ the chiral parameter $\s$ and its conjugate $\bar \s$ appear in  \eqref{SU(N)sW} only in the real combination $\s + \bar \s$.

In the case that $\cD_A = D_A = (\pa_a , D_\a^i , \bar D^\ad_i)$, the covariant derivatives of $\cN$-extended Minkowski superspace ${\mathbb M}^{4|4\cN}$, the relations \eqref{SU(N)sW}
provide a conformally flat realisation for an arbitrary conformally flat superspace.


\subsection{Degauging (iii): $\cN$-extended AdS superspace}

As an application of the superspace geometries sketched above, we now show how the $\cN$-extended AdS supergeometry may be described within $\sSU(\cN)$ superspace.
Such a supergeometry is characterised by the following conditions:

(i) the torsion and curvature tensors are Lorentz invariant;

(ii) the torsion and curvature tensors are covariantly constant.

These conditions imply the following relations:
\begin{subequations}
\begin{align}
	\cN=1:& \qquad G_\aa = 0~, \qquad \cD_A R = 0~. \\
	\label{6.26} \cN>1:& \qquad Y_{\a \b}^{ij} = 0 ~, \qquad G_\aa = 0~, \qquad \cD_A S^{jk} = 0~.
\end{align}
\end{subequations}
Keeping in mind these constraints, the algebra obeyed by ${\cD}_A$ reduces to the following:
\begin{subequations} 
	\label{2.170}
	\bea
	\{ \cD_\a^i , \cD_\b^j \}
	&=&
	4 S^{ij}  M_{\a\b} 	- 4 \ve_{\a\b}S^{k[i}  \mathbb{J}^{j]}{}_{k}
	~,
	\\
	\{ \bar{\cD}^\ad_i , \bar{\cD}^\bd_j \}
	&=&
	- 4 \bar{S}_{ij}  \bar{M}_{\ad\bd} 	+ 4 \ve^{\ad \bd} \bar{S}_{k[i}  \mathbb{J}^{k}{}_{j]}
	~,
	\\
	\{ \cD_\a^i , \bar{\cD}^\bd_j \}
	&=&
	- 2 \ri \d_j^i\cD_\a{}^\bd
	~,
	\\
	{[} \cD_\a^i, \cD_{\bb}{]}&=& 
	- \ri \ve_{\a \b} S^{ij} \bar{\cD}_{\bd j}
	~,
	\qquad
	{[} \bar{\cD}^\ad_i, \cD_{\bb}{]}= 
	\ri \d_\bd^\ad \bar{S}_{ij} \cD_\b^j
	~, 
	\\
	\big [ \cD_\aa , \cD_\bb \big] &=& - \frac{2}{\cN} S^{ij} \bar{S}_{ij} (\ve_{\a \b} \bar{M}_{\ad \bd} + \ve_{\ad \bd} M_{\a \b})~,
	\eea
\end{subequations}
with the identification $\bar{R} = - S$ when $\cN=1$. Additionally, in the $\cN=2$ case, one may impose the reality condition $\overline{S^{ij}} = S_{ij}$ by performing some rigid $\sU(1)$ phase transformation $\cD_\a^i \rightarrow \re^{\ri \phi} \cD_\a^i$. 
Then, the resulting geometry coincides with the one of \cite{KT-M08}. We will not impose this reality condition below.

When $\cN>1$, the constraint $\cD_A S^{jk} = 0$ implies the following integrability condition\footnote{In the $\cN=2$ case, a solution to eq. \eqref{6.42} is the reality condition $\overline{S^{ij}} = S_{ij}$.}
\begin{align}
	\label{6.42}
	\d_{(k}^{[i} S^{j]m} \bar{S}_{l)m} = 0 \quad \implies \quad 
	S^{ik} \bar{S}_{jk} = \frac{1}{\cN} \d^i_j S^{kl} \bar{S}_{kl}
	~.
\end{align}
This means that, by performing a local $\sU(\cN)_R$ transformation,\footnote{Strictly speaking, this should be performed within the $\sU(\cN)$ superspace of section \ref{Section6.2}, though it is also sufficient to introduce a flat $\sU(1)_R$ connection.} one can bring $S^{ij}$ to the form 
\begin{align}
	\label{6.43}
	S^{ij} = \d^{ij} S ~, \qquad S \in \mathbb{C} - \{ 0 \}~,
\end{align}
though it should be emphasised that our frame will no longer be conformally flat. As $\d^{ij}$ is the $\sSO(\cN)$ invariant tensor, it follows that the $R$-symmetry group reduces to $\sSO(\cN)_R$. The former may then be utilised to raise and lower indices in accordance with the rule
\begin{align}
	\psi^i = \d^{ij} \psi_j ~, \qquad \psi_i = \d_{ij} \psi^j~.
\end{align}
Further, upon inspection of \eqref{2.170}, the $R$-symmetry generators only appear in the algebra of covariant derivatives via the combination
\begin{align}
	\mathcal{J}^{ij} := - 2 \d^{k[i} \mathbb{J}^{j]}{}_k = - \cJ^{ji} 
	~, \qquad \cJ_{ij} = \d_{i k} \d_{j l} \cJ^{k l} = - \cJ_{j i}~.
\end{align}
The $\sSO(\cN)_R$ generator $\cJ^{ij}$ may be shown to act on isospinors as follows
\begin{align}
	\cJ^{ij} \psi^k = 2 \d^{k[i} \psi^{j]}~.
\end{align}
The resulting algebra of covariant derivatives is as follows:
\begin{subequations} 
	\label{AdS}
	\bea
	\{ \cD_\a^i , \cD_\b^j \}
	&=&
	4 S \d^{ij}  M_{\a\b} + 2 \ve_{\a\b} S \cJ^{ij}
	~,
	\\
	\{ \bar{\cD}^\ad_i , \bar{\cD}^\bd_j \}
	&=&
	- 4 \bar{S} \d_{ij}  \bar{M}^{\ad\bd} - 2 \ve^{\a\b} \bar{S}  \cJ_{ij}
	~,
	\\
	\{ \cD_\a^i , \bar{\cD}^\bd_j \}
	&=&
	- 2 \ri \d_j^i\cD_\a{}^\bd
	~,
	\\
	{[} \cD_\a^i, \cD_{\bb}{]}&=& 
	- \ri \ve_{\a \b} S \bar{\cD}_{\bd}^i
	~,
	\qquad
	{[} \bar{\cD}^\ad_i, \cD_{\bb}{]}= 
	\ri \d_\bd^\ad \bar{S} \cD_{\b j}
	~, 
	\\
	\big [ \cD_\aa , \cD_\bb \big] &=& - 2 S \bar{S} (\ve_{\a \b} \bar{M}_{\ad \bd} + \ve_{\ad \bd} M_{\a \b})~.
	\eea
\end{subequations}
This algebra coincides with the one presented in eq. \eqref{NowarAlgebra} provided one fixes $S = -2$, which indicates that, for $\cN>1$, the latter also does not describe a conformally flat frame. This will be elaborated on in a forthcoming work.

We now relax the constraint \eqref{6.43} and provide a manifestly conformally flat realisation of AdS superspace. By definition, a conformally flat supergeometry may be related to a flat one by performing some super-Weyl transformation. In the case of AdS superspace, this means that the curved covariant derivatives $\cD_A$ are related to those of Minkowski superspace $D_A = (\pa_a,D_\a^i,\bar{D}^\ad_i)$, see eq. \eqref{SU(N)sW}, as follows:
\begin{subequations}
	\label{AdSBoost}
	\begin{align}
		\cD_\a^i &= \re^{\frac{\cN-2}{2\cN} \s + \frac 1 \cN \bar{\s}} \Big( D_\a^i+ D^{\b i}\s M_{\a \b} + D_{\a}^j \s \mathbb{J}^{i}{}_j \Big) ~, \\
		\bar{\cD}_{i}^{' \ad}&=\re^{\frac{1}{\cN} \s + \frac{\cN-2}{2\cN} \bar{\s}} \Big( \bar{D}^\ad_i- \bar{D}_{ \bd i} \bar{\s} \bar{M}^{\ad \bd} - \bar{D}^{\ad}_j \bar{\s} \mathbb{J}^{j}{}_i \Big)~,
		\\
		\cD_\aa &= \re^{\hf \s + \hf \bar{\s}} \Big(\partial_\aa + \frac{\rm i}{2} D^i_{\a} \s \bar{D}_{\ad i} + \frac{\rm i}{2} \bar{D}_{\ad i} \bar{\s} D_{\a}^i + \hf \Big( \partial^\b{}_\ad (\s + \bar \s ) - \frac{\ri}{2} D^{\b i} \s \bar{D}_{\ad i} \bar{\s} \Big) M_{\a \b} \non \\ & \qquad \qquad \quad + \hf \Big( \partial_{\a}{}^\bd (\s + \bar{\s}) + \frac{\ri}{2} D_{\a}^i \s \bar{D}^{\bd}_i \bar{\s} \Big) { \bar M}_{\ad \bd} \Big)~,
	\end{align}
while the AdS superspace curvature $S^{ij}$ takes the form
	\begin{align}
		S^{ij }= - \frac 1 4 \re^{\frac{\cN-2}{\cN} \s + \frac 2 \cN \bar{\s}} \Big(D^{ij} \s - D^{\a i} \s D_\a^{j} \s \Big)~.
	\end{align}
Here the chiral parameter $\s$ is required to obey the following constraints:
	\begin{align}
	D_{(\a}^{[i} D_{\b)}^{j]} \re^\s = 0~, \qquad [D_\a^i,\bar{D}_{\ad i}] \re^{\frac{\cN}{2}(\s + \bar{\s})} = 0 
	~.
	\end{align}
\end{subequations}
As compared with \cite{BILS}, our work provides an alternative proof of the conformal flatness of $\cN$-extended AdS superspace. It should also be pointed out that the logarithm of the chiral parameter $\l$, which was defined in equations \eqref{chiral chart} and \eqref{chiral chart2}, is proportional to $\s$; $\text{ln}\,\l \propto \s$. Further, in $\cN=1$ case, they are related via eq. \eqref{5.37}.


\section{Conclusion}

This work has completed the construction of the embedding formalism for $\text{AdS}^{4|4} $ initiated in \cite{KT-M21}. In the original realisation \cite{KT-M21}, superspace Poincar\'e coordinates for $\text{AdS}^{4|4\cN}$ are naturally introduced, and therefore that realisation is well suited for AdS/CFT calculations in the spirit of \cite{BFP}.
The novel realisation of the ${\cal N}$-extended AdS supergroup $\mathsf{OSp}(\mathcal{N}|4;\mathbb{R})$, which has been introduced in this paper, is more suitable for the coset construction, 

The AdS superparticle model \eqref{superparticle model} is one of the main results of this paper. Setting $\k=0$ in \eqref{superparticle model}  gives a unique AdS extension of the model for a massive superparticle in Minkowski superspace. In terms of the local coordinates in the north chart described in subsection \ref{n s cosets}, the kinetic terms have the form 
\bea
	{\rm Str}(\dot{\bar{{X}}}\dot{{X}}) = -4\ell^2 \l^{2}\bar{\l}^{2} \eta_{mn}\dot{e}^{m}\dot{e}^{n}
= -4 \eta_{mn}\dot{e}^{m}\dot{e}^{n} + \cO(\ell^{-1}) ~,
\eea
where the one-form $e^m$ is defined in \eqref{4.27}.
In the non-supersymmetric case, $\N=0$, the $\k$-term is absent. 
Therefore, for $\cN>0$ the $\k$-term does not contain purely bosonic contributions. 
It may be checked that the $\k$-term contains a higher-derivative contribution 
proportional to $ \ri \ell ( \dot{\q}^2 - \dot{\bar \q}^2 )$.
Thus our superparticle model \eqref{superparticle model} may be viewed as  an AdS analogue of the Volkov-Pashnev model \cite{Volkov:1980mg}.\footnote{We are grateful to Dmitri Sorokin for bringing the references
\cite{Volkov:1980mg, deAzcarraga:1982dhu}
 to our attention.} In $\cN$-extended Minkowski superspace, for $\cN>1$ it was possible to add a fermionic  WZ-like term to the superparticle action \cite{deAzcarraga:1982dhu}. Such structures are more difficult to generate in the AdS case. 


In this paper we have also provided descriptions of the most general conformally flat $\cN$-extended supergeometry in four dimensions. Specifically, we have realised this geometry in three different superspace frameworks: (i) conformal superspace; (ii) $\sU(\cN)$ superspace; and (iii) $\sSU(\cN)$ superspace. Additionally, we computed the finite super-Weyl transformations within the $\sU(\cN)$ and $\sSU(\cN)$ superspaces. As an application of this construction, we utilised it to obtain a new realisation for $\text{AdS}^{4|4\cN}$ and describe the specific super-Weyl transformation \eqref{AdSBoost} required to `boost' to this superspace from a flat one.

\noindent
{\bf Acknowledgements:}\\
We are grateful to Alex Arvanitakis, Dmitri Sorokin and Gabriele Tartaglino-Mazzucchelli for discussions. SMK is grateful to the organisers of the CQUeST-APCTP Workshop ``Gravity beyond Riemannian Paradigm'' (Jeju Island, South Korea) 
where part of this work was completed, for the fantastic scientific atmosphere and generous support. He also acknowledges kind hospitality and generous support extended to him during his research stay at KIAS, Seoul. The work of SMK and ESNR is supported in part by the Australian Research Council, projects DP200101944 and DP230101629.
The work of NEK is supported by the Australian Government Research Training Program Scholarship.

\appendix


\section{The supergroup $\sOSp(\mathcal{N}|4;\mathbb{R})$ and supertwistors} \label{Supertwistors}

In this appendix we collect essential definitions concerning the supergroup  $\sOSp(\mathcal{N}|4;\mathbb{R})$ which has two different but related origins in supersymmetric field theory: 
(i) as
 the $\cN$-extended superconformal group in three dimensions; and 
(ii) as
 the $\cN$-extended AdS supergroup in four dimensions.  We start by discussing the complex supergroup  $\sOSp(\mathcal{N}|4;\mathbb{C})$ of which  $\sOSp(\mathcal{N}|4;\mathbb{R})$ is a real form.

The supergroup 
$\sOSp(\mathcal{N}|4;\mathbb{C})$ 
naturally acts on the space of {\it  even} supertwistors and on
the space of {\it  odd} supertwistors.
An arbitrary supertwistor is a column vector
\bea
T = (T_A) =\left(
\begin{array}{c}
T_\hal \\
\hline \hline
 T_I
\end{array}
\right)~, 
\qquad \hat \a = 1,2, 3,4 ~, \quad I = 1, \dots, \cN ~.
\eea
In the case of even supertwistors, $ T_\hal$ is bosonic
and $T_I$ is fermionic.
In the case of odd supertwistors, $ T_\hal$ is fermionic while  $T_I$ is bosonic.
The even and odd supertwistors are called pure.
It is useful to introduce the parity function $\e ( T )$ defined as:
$\e ( T ) = 0$ if $ T$ is even, and $\e ( T ) =1$ if $T $ is odd.
It is also useful to define
\bea
 \e_A = \left\{
\begin{array}{c}
 0 \qquad A=\hal \\
 1 \qquad A=I
\end{array}
\right.{}~.
\non
\eea
Then the components $T_A$ of a pure supertwistor
 have the following  Grassmann parities
\bea
\e ( T_A) = \e ( T ) + \e_A \quad (\mbox{mod 2})~.
\eea
The space of  even supertwistors is naturally identified with
${\mathbb C}^{4|\cN}$,
while the space of  odd supertwistors may be identified with
${\mathbb C}^{\cN |4}$.

Let us introduce the following  graded antisymmetric $(4|\cN) \times (4|\cN)$
supermatrix
\bea
{\mathbb J} = ({\mathbb J}^{AB}) = \left(
\begin{array}{c ||c}
J ~&~ 0 \\\hline \hline
0 ~& ~{\rm i} \,{\mathbbm 1}_\cN
\end{array} \right) ~, \qquad
J
=\big(J^{\hat \a \hat \b} \big)
=\left(
\begin{array}{cc}
0  & {\mathbbm 1}_2\\
 -{\mathbbm 1}_2  &    0
\end{array}
\right) ~,
\label{supermetric}
\eea
where ${\mathbbm 1}_n $ denotes the unit  $n \times n$ matrix.
Making use of $\mathbb J$ allows us to define a graded symplectic inner product on the space of pure supertwistors by the rule: for arbitrary pure supertwistors $T$ and $S$,
the inner product is
\bea
\langle T | S \rangle  : = T^{\rm sT} {\mathbb J} \, S
~,
\label{innerp}
\eea
where the row vector  $T^{\rm sT} $ is defined by
\bea
{ T}^{\rm sT} := \big( T_\hal , - (-1)^{\e(T)}  T_I \big)
= (  T_A (-1)^{\e(T)\e_A +\e_A} )
\eea
and is called the super-transpose of $T$.
The above inner product is graded anti-symmetric:
\bea
\langle T_1 |  T_2  \rangle
= -(-1)^{ \e (T_1) \e(T_2) }  \langle T_2 | T_1  \rangle ~.
\eea

The supergroup   $\sOSp(\cN|4; {\mathbb C})$ is defined to 
consist of those even $(4|\cN) \times (4|\cN)$ supermatrices
\bea
g = (g_A{}^B) ~, \qquad \e(g_A{}^B) = \e_A + \e_B ~,
\eea
which preserve the inner product \eqref{innerp} under the action
\bea
T =(T_A) ~\to ~ gT = (g_A{}^B T_B)~. \label{2.9}
\eea
Such a transformation maps the space of even (odd) supertwistors onto itself.
The condition of invariance of the inner  product \eqref{innerp}
under \eqref{2.9} is
\bea
g^{\rm sT} {\mathbb J}\, g = {\mathbb J} ~, \qquad
(g^{\rm sT})^A{}_B := (-1)^{\e_A \e_B + \e_B} g_B{}^A~.
\label{groupcond}
\eea
It is useful to recast the definition of $g^{\rm sT}$
in an equivalent form by writing $g$ as a block supermatrix: 
\begin{align} \label{originalosp}
	g =  \left(
	\begin{array}{c||c}
		A & B\\
		\hline \hline
		C & D
	\end{array}
	\right)\,,
	\qquad g^{\sT} =\left(
	\begin{array}{c||c}
		A^\T & -C^\T\\
		\hline \hline
		B^{\T} & D^\T
	\end{array}
	\right)\,.
\end{align}

A pure supertwistor is said to be real if its components obey the reality condition
\bea
\overline{T_A} = (-1)^{\e(T) \e_A + \e_A} T_A~.
\label{initrealitycond}
\eea
The space of real even supertwistors is naturally identified with
${\mathbb R}^{4|\cN}$,
while the space of real odd supertwistors may be identified with
${\mathbb R}^{\cN |4}$.
Given two real supertwistors $T$ and $S$, 
it holds that
\bea
\overline{
\langle { T}| { S} \rangle }=
-\langle {S}| { T} \rangle ~.
\eea
The reality condition \eqref{initrealitycond} is not preserved under the action \eqref{2.9} of 
$ \sOSp(\cN|4; {\mathbb C})$.

By definition, the real subgroup $\sOSp(\cN|4; {\mathbb R}) \subset \sOSp(\cN|4; {\mathbb C})$
consists of those transformations which preserve the reality condition
\eqref{initrealitycond},
\begin{subequations}
\bea
\overline{T_A} = (-1)^{\e(T) \e_A + \e_A} T_A \quad \longrightarrow \quad
\overline{(gT)_A} = (-1)^{\e(T) \e_A + \e_A} (gT)_A~.
\eea
This is equivalent  to
\bea
\overline{ g_A{}^B} = (-1)^{\e_A \e_B + \e_A} g_A{}^B \quad
\Longleftrightarrow \quad g^\dagger = g^{\rm sT}~, \qquad
\forall g \in \sOSp(\cN|4; {\mathbb R}) ~.
\label{A.13b}
\eea
In conjunction with \eqref{groupcond}, this reality condition can be recast in the form
\bea
g^\dagger {\mathbb J}\, g = {\mathbb J} ~.
\label{A.13c}
\eea
\end{subequations}

In the case of complex supertwistors, the following involution can be defined 
\bea
T = (T_A) ~ \to ~*T =\big( (* T)_A\big) ~, \qquad (* T)_A:=  (-1)^{\e(T) \e_A + \e_A} \overline{T_A}~, \qquad *(*T)=T~.
\label{A.14}
\eea
Its crucial property is that $*T$ is a supertwistor with respect to 
$\sOSp(\cN|4; {\mathbb R})$,
\bea
g (*T )= *(gT)~, \qquad \forall g \in \sOSp(\cN|4; {\mathbb R})~.
\eea
We also observe that 
\bea
(*T)^{\rm sT} = T^\dagger~.
\eea
Given a real supertwistor $T$ satisfying the reality condition \eqref{initrealitycond}, it holds that  $*T =T$. 


\section{Conformally flat atlas for AdS${}_d$} \label{Stereographic}

A $d$-dimensional AdS space,  AdS$_d$,
can be identified with a hypersurface in pseudo-Euclidean space ${\mathbb R}^{d-1,2}$  
defined by 
  \bea
 -(Z^0)^2 + (Z^1)^2 + \dots +(Z^{d-1})^2 - (Z^d)^2 
 \equiv \eta_{ab} Z^a Z^b - (Z^d)^2 
 = -\ell^2 ={\rm const}~,
 \label{Embedding}
 \eea
 where $Z^A \equiv(Z^a, Z^d)$ denotes the Cartesian coordinates of ${\mathbb R}^{d-1,2}$,
 with   $A = 0, 1, \dots, d-1, d $ and $a = 0, 1, \dots, d-1$.
One can cover AdS$_d$ by two charts: 
(i) the north chart in which 
$Z^d> -\ell$; and 
(ii) the south chart in which $Z^d<\ell$.\footnote{The north chart and the south chart are everywhere dense open subsets of AdS$_d$. In particular, those points of AdS$_d$, which do not belong to the north chart, are characterised by the conditions $Z^d =-\ell$ and $\eta_{ab} Z^a Z^b =0$, and therefore they span a light cone in ${\mathbb R}^{d-1,1}$. }
Each chart may be parametrised using  a natural  generalisation of the stereographic projection for a $d$-dimensional sphere $S^d$.  
Given a point $Z^A =(Z^a, Z^d) $ in the north chart, its local coordinates $x^a$ will be chosen to correspond to the intersection of the hyperplane $Z^d=0$ 
with the straight line $\G^A_{\mathfrak N} (t)$ connecting $Z^{A}$ and the ``north pole'' 
$Z^{A}_{\rm north} = (0, \dots , 0, -\ell)$. 
Similarly, given a point $Z^A$ in the south chart, its local coordinates $y^a$ will be 
chosen to correspond to the intersection of the hyperplane $Z^d=0$ 
with  the straight line $\G^A_{\mathfrak S} (t)$ connecting $Z^A$ and the ``south pole'' 
$Z^A_{\rm south} = (0, \dots , 0, \ell)$. 

In the north chart, the straight line $\G^A_{\mathfrak N} (t)$ can be parametrised 
as 
\bea
\G^A_{\mathfrak N} (t) = (1-t) Z^{A}_{\rm north} + t x^A~, \qquad x^A = (x^a, 0)~,
\eea
and $Z^A \in {\rm AdS}_d$ corresponds to some value $t'$ of the evolution parameter, $\G^A_{\mathfrak N} (t') = Z^A$. We then derive
\bea
x^a = \frac{\ell Z^a}{\ell +Z^d} ~ \implies ~
x^2 := \eta_{ab}x^a x^b = - \ell^2 \frac{\ell - Z^d}{\ell + Z^d}
~ \implies ~ \ell^2 - x^2 = \frac{2\ell^3}{\ell + Z^d} >0~.
\eea
The embedding coordinates $Z^A$ can be expressed in terms of the local ones, 
\bea
Z^a = \frac{2\ell^2}{\ell^2 - x^2} x^a~, \qquad Z^d = \ell \frac{\ell^2 +x^2} {\ell^2 - x^2}~.
\eea
For the induced metric we obtain 
\bea
\rd s^2_{\rm AdS} = \frac{4\ell^4}{(\ell^2 - x^2)^2} \eta_{ab} \rd x^a \rd x^b~.
\eea

In the south chart, the straight line $\G^A_{\mathfrak S} (t)$ can be parametrised 
as 
\bea
\G^A_{\mathfrak S} (t) = (1-t) Z^{A}_{\rm south} + t y^A~, \qquad y^A = (y^a, 0)~,
\eea
and $Z^A \in {\rm AdS}_d$ corresponds to some value $t''$ of the evolution parameter, $\G^A_{\mathfrak S} (t'') = Z^A$. We obtain
\bea
y^a = \frac{\ell Z^a}{\ell -Z^d} ~ \implies ~
y^2 := \eta_{ab}y^a y^b = - \ell^2 \frac{\ell + Z^d}{\ell - Z^d}
~ \implies ~ \ell^2 - y^2 = \frac{2\ell^3}{\ell - Z^d} >0~.
\eea
The embedding coordinates $Z^A$ are expressed in terms of the local ones $y^a$ as follows: 
\bea
Z^a = \frac{2\ell^2}{\ell^2 - y^2} y^a~, \qquad Z^d = -\ell \frac{\ell^2 +y^2} {\ell^2 - x^2}~.
\eea
The induced metric has the form
\bea
\rd s^2_{\rm AdS} = \frac{4\ell^4}{(\ell^2 - y^2)^2} \eta_{ab} \rd y^a \rd y^b~.
\eea

It remains to consider the intersection of the north and south charts, which is characterised by 
$-\ell <Z^d < \ell$. A short calculation of the transition functions gives
\bea
y^a = - \frac{\ell^2}{x^2} x^a~~\implies ~ ~y^2 x^2 = \ell^4~. \label{B.9}
\eea
It may also be seen that $x^2 <0 \Longleftrightarrow y^2 <0$ in the intersection of the charts.


\section{The $\cN$-extended superconformal algebra} \label{AppendixC}

In this appendix, we spell out our conventions for the $\cN$-extended superconformal algebra of Minkowski superspace, $\mathfrak{su}(2,2|\cN)$. It was initially described in the literature by Park \cite{Park}, see also \cite{KTh}. We emphasise that the appropriate relations differ by an overall sign as compared with those of eq. \eqref{superalgebra}. This distinction arises from our adoption of the convention where generators act on fields and operators in a consistent manner. 

The conformal algebra, $\mathfrak{su}(2,2)$, consists of the translation $(P_a)$, Lorentz $(M_{ab})$, special conformal $(K_a)$ and dilatation $(\mathbb{D})$ generators. Amongst themselves, they obey the algebra
\begin{subequations} 
	\label{2.17}
	\begin{align}
		&[M_{ab},M_{cd}]=2\eta_{c[a}M_{b]d}-2\eta_{d[a}M_{b]c}~, \phantom{inserting blank space inserting} \\
		&[M_{ab},P_c]=2\eta_{c[a}P_{b]}~, \qquad \qquad \qquad \qquad ~ [\mathbb{D},P_a]=P_a~,\\
		&[M_{ab},K_c]=2\eta_{c[a}K_{b]}~, \qquad \qquad \qquad \qquad [\mathbb{D},K_a]=-K_a~,\\
		&[K_a,P_b]=2\eta_{ab}\mathbb{D}+2M_{ab}~.
	\end{align}
\end{subequations}

The $R$-symmetry group $\sU(\cN)_R$ is generated by the $\sU(1)_R$ $(\mathbb{Y})$ and $\sSU(\cN)_R$ $(\mathbb{J}^i{}_j)$ generators, which commute with all elements of the conformal algebra. Amongst themselves, they obey the commutation relations
\begin{align}
	[\mathbb{J}^{i}{}_j,\mathbb{J}^{k}{}_l] = \d^i_l \mathbb{J}^k{}_j - \d^k_j \mathbb{J}^i{}_l ~.
\end{align}

The superconformal algebra is then obtained by extending the translation generator to $P_A=(P_a,Q_\a^i,\bar{Q}^\ad_i)$ and the special conformal generator to $K^A=(K^a,S^\a_i,\bar{S}_\ad^i)$. The commutation relations involving the $Q$-supersymmetry generators with the bosonic ones are:
\begin{subequations} 
	\bea
	\big[M_{ab}, Q_\g^i \big] &=& (\s_{ab})_\g{}^\d Q_\d^i ~,\quad 
	\big[M_{ab}, \bar Q^\gd_i \big] = (\tilde{\s}_{ab})^\gd{}_\dd \bar Q^\dd_i~,\\
	\big[\mathbb{D}, Q_\a^i \big] &=& \hf Q_\a^i ~, \quad
	\big[\mathbb{D}, \bar Q^\ad_i \big] = \hf \bar Q^\ad_i ~, \\
	\big[\mathbb{Y}, Q_\a^i \big] &=&  Q_\a^i ~, \quad
	\big[\mathbb{Y}, \bar Q^\ad_i \big] = - \bar Q^\ad_i ~, \label{2.19c} \\
	\big[\mathbb{J}^i{}_j, Q_\a^k \big] &=&  - \d^k_j Q_\a^i + \frac{1}{\mathcal{N}} \d^i_j Q_\a^k ~, \quad
	\big[\mathbb{J}^i{}_j, \bar Q^\ad_k \big] = \d^i_k \bar Q^\ad_j - \frac{1}{\mathcal N} \d^i_j \bar Q^\ad_k ~,  \\
	\big[K^a, Q_\b^i \big] &=& -\ri (\s^a)_\b{}^\bd \bar{S}_\bd^i ~, \quad 
	\big[K^a, \bar{Q}^\bd_i \big] = 
	-\ri ({\s}^a)^\bd{}_\b S^\b_i ~.
	\eea
\end{subequations}
At the same time, the commutation relations involving the $S$-supersymmetry generators 
with the bosonic operators are: 
\begin{subequations}
	\bea
	\big [M_{ab} , S^\g_i \big] &=& - (\s_{ab})_\b{}^\g S^\b_i ~, \quad
	\big[M_{ab} , \bar S_\gd^i \big] = - (\ts_{ab})^\bd{}_\gd \bar S_\bd^i~, \\
	\big[\mathbb{D}, S^\a_i \big] &=& -\hf S^\a_i ~, \quad
	\big[\mathbb{D}, \bar S_\ad^i \big] = -\hf \bar S_\ad^i ~, \\
	\big[\mathbb{Y}, S^\a_i \big] &=&  -S^\a_i ~, \quad
	\big[\mathbb{Y}, \bar S_\ad^i \big] =  \bar S_\ad^i ~,  \label{2.20c}\\
	\big[\mathbb{J}^i{}_j, S^\a_k \big] &=&  \d^i_k S^\a_j - \frac{1}{\mathcal{N}} \d^i_j S^\a_k ~, \quad
	\big[\mathbb{J}^i{}_j, \bar S_\ad^k \big] = - \d_j^k \bar S_\ad^i + \frac{1}{\mathcal N} \d^i_j \bar S_\ad^k ~,  \\
	\big[ S^\a_i , P_b \big] &=& \ri (\s_b)^\a{}_\bd \bar{Q}^\bd_i ~, \quad 
	\big[\bar{S}_\ad^i , P_b \big] = 
	\ri ({\s}_b)_\ad{}^\b Q_\b^i ~.
	\eea
\end{subequations}
Finally, the anti-commutation relations of the fermionic generators are: 
\begin{subequations}
	\bea
	\{Q_\a^i , \bar{Q}^\ad_j \} &=& - 2 \ri \d^i_j (\s^b)_\a{}^\ad P_b=- 2 \ri \d^i_j  P_\a{}^\ad~, \\
	\{ S^\a_i , \bar{S}_\ad^j \} &=& 2 \ri  \d_i^j (\s^b)^\a{}_\ad K_b=2 \ri \d_i^j  K^\a{}_\ad
	~, \\
	\{ S^\a_i , Q_\b^j \} &=& \d_i^j \d^\a_\b \Big(2 \mathbb{D} + \frac{\mathcal{N}-4}{\mathcal{N}}\mathbb{Y} \Big) - 4 \d_i^j  M^\a{}_\b 
	+ 4 \d^\a_\b  \mathbb{J}^j{}_i ~, \\
	\{ \bar{S}_\ad^i , \bar{Q}^\bd_j \} &=& \d_j^i \d^\bd_\ad \Big(2 \mathbb{D} - \frac{\mathcal{N}-4}{\mathcal{N}}\mathbb{Y} \Big) + 4 \d_j^i  \bar{M}_\ad{}^\bd 
	- 4 \d_\ad^\bd  \mathbb{J}^i{}_j  ~. \label{2.21d}
	\eea
\end{subequations}
We emphasise that all (anti-)commutators not listed above vanish identically.


\begin{footnotesize}

\end{footnotesize}

\end{document}